\documentclass[lettersize,journal]{IEEEtran}
\usepackage{amsmath,amsfonts}
\usepackage{algorithmic}
\usepackage{algorithm}
\usepackage{array}
\usepackage[caption=false,font=normalsize,labelfont=sf,textfont=sf]{subfig}
\usepackage{textcomp}
\usepackage{stfloats}
\usepackage{url}
\usepackage{verbatim}
\usepackage{graphicx}
\usepackage{cite}
\usepackage{acronym}
\usepackage{cleveref}
\usepackage{amsmath,amssymb,amsfonts}
\usepackage{mathtools}
\crefname{figure}{Fig.}{Figs.}%
\usepackage{multirow}
\usepackage{makecell}
\usepackage{amsmath,amssymb,amsfonts}
\usepackage{mathtools}
\usepackage{svg}
\def\BState{\State\hskip-\ALG@thistlm}
\makeatother

\DeclareMathAlphabet{\mathbit}{OML}{cmr}{bx}{it}
\newcommand{\B}[1]{\mathbf{#1}}

\usepackage{tikz}
\usetikzlibrary{calc}

\newcommand{\positiontextbox}[4][]{%
	\begin{tikzpicture}[remember picture,overlay]
		\node[inner sep=3pt, fill=yellow,align=left,draw,line width=1pt,#1] at ($(current page.north west) + (#2,-#3)$) {\parbox{.95\paperwidth}{#4}};
	\end{tikzpicture}%
}

\def\B{\mathbf}

\acrodef{5G}{fifth generation}
\acrodef{4G}{fourth generation}
\acrodef{eMBB}{Enhanced  Mobile  Broadband}
\acrodef{LTE}{Long Term Evolution}
\acrodef{EM}{electromagnetic}
\acrodef{URLLC}{Ultra-Reliable  and  Low Latency  Communications}
\acrodef{mMTC}{Massive Machine-Type Communications}
\acrodef{CIoT}{Cellular Internet  of  Things}
\acrodef{V2X}{Vehicle-to-Everything}
\acrodef{MIMO}{multiple-input multiple-output}
\acrodef{MISO}{multiple-input single-output}
\acrodef{B5G}{beyond 5G}
\acrodef{IRS}{intelligent reflecting surface}
\acrodef{LoS}{line-of-sight}
\acrodef{MU-MIMO}{multi-user MIMO}
\acrodef{EE}{energy efficiency}
\acrodef{SE}{spectrum efficiency}
\acrodef{LoS}{line-of-sight}
\acrodef{NLoS}{non-line-of-sight}
\acrodef{BS}{base station}
\acrodef{MEC}{mobile edge computing}
\acrodef{IoT}{Internet of Things}
\acrodef{6G}{sixth generation}
\acrodef{SNR}{signal-to-noise ratio}
\acrodef{DL}{deep learning}
\acrodef{ANN}{artificial neural network}
\acrodef{DRL}{deep reinforcement learning}
\acrodef{RL}{reinforcement learning}
\acrodef{DDPG}{deep deterministic policy gradient}
\acrodef{MS}{multi-stream}
\acrodef{DQL}{Deep Q-learning}
\acrodef{PG}{policy gradient} 
\acrodef{SU-MISO}{single-user MISO}
\acrodef{MU-MISO}{multi-user multiple-input single-output}
\acrodef{CSI}{channel state information}
\acrodef{RF}{radio frequency} 
\acrodef{AWGN}{additive white Gaussian noise}
\acrodef{MMSE}{minimum mean square error}
\acrodef{DPG}{deterministic policy gradient}
\acrodef{ReLU}{rectified linear unit}
\acrodef{tanh}{hyperbolic tangent}
\acrodef{SS}{single-stream}
\acrodef{DoF}{degrees of freedom} 
\acrodef{UAV}{unmanned aerial vehicle}
\acrodef{MRT}{maximum ratio transmitter}
\acrodef{TD3PG}{twin-delayed DDPG}
\acrodef{CB}{contextual bandit}
\acrodef{DCB}{deep CB}
\acrodef{DCB-DDPG}{deep contextual bandit-oriented deep deterministic policy gradient}
\acrodef{MDP}{Markov decision process}
\acrodef{mmWave}{mm-wave}
\acrodef{PPO}{proximal policy optimization}
\acrodef{TRPO}{trust region policy optimization}
\acrodef{A3C}{asynchronous advantage actor-critic}
\acrodef{POMDP}{partially observable MDP}

\begin{document}

\onecolumn
\begingroup

\setlength\parindent{0pt}
\fontsize{14}{14}\selectfont

\vspace{1cm} 
\textbf{This is an ACCEPTED VERSION of the following published document:}

\vspace{1cm} 
D. Pereira-Ruisánchez, O. Fresnedo, D. Pérez-Adán and L. Castedo,``Deep Contextual Bandit and Reinforcement Learning for IRS-Assisted MU-MIMO Systems'', \textit{IEEE Transactions on Vehicular Technology}, vol. 72, n.o 7, pp. 9099-9114, Jul 2023, doi: 10.1109/TVT.2023.3249353

\vspace{1cm} 
Link to published version: https://doi.org/10.1109/TVT.2023.3249353

\vspace{3cm}

\textbf{General rights:}

\vspace{1cm} 
\textcopyright 2023 IEEE. This version of the article has been accepted for publication, after peer review. {Personal use of this material is permitted. Permission from IEEE must be obtained for all other uses, in any current or future media, including reprinting/republishing this material for advertising or promotional purposes, creating new collective works, for resale or redistribution to servers or lists, or reuse of any copyrighted component of this work in other works.}
\twocolumn
\endgroup
\clearpage

\title{Deep Contextual Bandit and Reinforcement Learning for IRS-assisted MU-MIMO Systems}

\author{Dariel Pereira-Ruisánchez,~\IEEEmembership{Student Member,~IEEE,} Óscar Fresnedo,~\IEEEmembership{Member,~IEEE,} \\Darian Pérez-Adán, and Luis Castedo,~\IEEEmembership{Senior Member,~IEEE}}

\maketitle

\begin{abstract}
The combination of \ac{MIMO} systems and \acp{IRS} is foreseen as a critical enabler of \ac{B5G} and 6G. In this work, two different approaches are considered for the joint optimization of the \ac{IRS} phase-shift matrix and MIMO precoders of an \ac{IRS}-assisted \ac{MS} \ac{MU-MIMO} system. Both approaches aim to maximize the system sum-rate for every channel realization. The first proposed solution is a novel \ac{CB} framework with continuous state and action spaces called \ac{DCB-DDPG}. The second is an innovative \ac{DRL} formulation where the states, actions, and rewards are selected such that the \ac{MDP} property of \ac{RL} is appropriately met. Both proposals perform remarkably better than state-of-the-art heuristic methods in scenarios with high multi-user interference.

\end{abstract}


\acresetall

\positiontextbox{10.8cm}{26.1cm}{\footnotesize \textcopyright 2023 IEEE. This version of the article has been accepted for publication, after peer review. Personal use of this material is permitted. Permission from IEEE must be obtained for all other uses, in any current or future media, including reprinting/republishing this material for advertising or promotional purposes, creating new collective works, for resale or redistribution to servers or lists, or reuse of any copyrighted component of this work in other works. Published version:
	https://doi.org/10.1109/TVT.2023.3249353}

\section{Introduction}
\IEEEPARstart{T}{he} research interest in \ac{IRS}-assisted \ac{MIMO} communication systems is enormous nowadays. \acp{IRS} are two-dimensional surfaces composed of large arrays of passive scattering elements with specially designed physical structures \cite{liang_large_2019}. Each scattering element can be individually controlled in a software-defined manner to change its \ac{EM} properties and, this way, the phases of the impinging signals. By properly designing the MIMO precoding/combining matrices and the IRS phase-shift matrix, the demanding requirements of the use cases in \ac{B5G} and \ac{6G} systems are expected to be accomplished \cite{gong_toward_2020, guan_joint_2020, castaneda_overview_2017, navarro-ortiz_survey_2020, zhao_massive_2018, perez-adan_intelligent_2021}.

In conventional MIMO systems, the communication channel is an uncontrollable factor with a high impact on the reception quality. In this sense, IRS-assisted communications are the most appealing alternative to configure the propagation environment smartly \cite{gong_toward_2020}. In \cite{huang_reconfigurable_2019}, the authors analyze the deployment of \ac{IRS}-assisted \ac{MU-MIMO} systems as an energy-efficient alternative to multi-antenna amplify-and-forward relaying. Numerical results show that the \ac{IRS}-based approach provides up to 300{\%} higher \ac{EE} while reaching near optimal \ac{SE}. Authors in \cite{guo_weighted_2020} address  the \ac{SE} performance of an IRS-assisted \ac{MU-MISO} downlink communication in a scenario with \ac{NLoS} conditions between the \ac{BS} and the users. As expected, the sum-rate values significantly increase when considering the \ac{BS}-IRS-user channel paths. 

Due to the dynamic nature of vehicular communications, \ac{IRS} is a promising technology for achieving cost and energy-efficient communications via smartly reshaping the wireless propagation channel \cite{al-hilo_reconfigurable_2022}. According to \cite{geraci_what_2021}, \ac{UAV} applications and cellular vehicular communications enhance by deploying \acp{IRS}. For instance, \acp{IRS} have been considered both to improve the reliability of the ground-to-air links and to be \ac{UAV}-carried to provide controllable on-demand coverage \cite{wei_sum-rate_2021}. 

Many challenges related to the deployment of \ac{IRS}-assisted MIMO systems remain unsolved \cite{chen_towards_2021, ma_joint_2020}. Control and deployment complexities are two major issues that increase with the number of configurable elements in both technologies. In \ac{IRS}-assisted systems, the average \ac{SNR} and the transmission power savings are directly related to the square of the number of scattering elements \cite{gong_toward_2020}. In \ac{MIMO} systems, the multiplexing gain, the spatial diversity, and the favorable propagation conditions improve when considering more antennas \cite{bjornson_massive_2017}. However, in \ac{MIMO} and \ac{IRS}-assisted communications, increasing the number of configurable elements leads to optimization problems that are often too computationally expensive and incur considerable timing overhead. 

On the one hand, the complexity of the optimization problems when considering \ac{IRS}-assisted communications increases because of some characteristics of the sets of feasible solutions. The passive behavior of the scattering elements in \acp{IRS} is usually modeled through complex-valued entries with modulus constraints, which are difficult to handle and lead to non-convex formulations of the problem. On the other hand, the requirement of jointly optimizing the users' precoders and the \ac{IRS} phase-shifts in the \ac{IRS}-assisted \ac{MIMO} systems leads to optimization problems with a non-convex cost function over a non-convex search space. Some conventional model-based approaches relax the original problem to obtain a convex formulation, approximate the non-convex constraints iteratively, develop heuristic solutions or change the optimization domain \cite{castaneda_overview_2017, hong_signal_2012, wang_beamforming_2021, ur_rehman_joint_2021, chang_capacity_2020, zhang_irs-aided_2020, tuan_secrecy_2020, huang_decentralized_2021}. However, in those cases, optimality is sacrificed for tractability. In addition, due to the increase in complexity and heterogeneity of these communications, the solutions to the optimization problems cannot wholly rely on models, which are either too computationally demanding or inconsistent with the actual behavior of practical deployments. Among others, these issues render conventional model-based approaches inefficient for most emerging time-sensitive applications.   

Several other approaches consider data-driven alternatives to deal with the high complexity of designing \ac{IRS}-assisted \ac{MIMO} systems. Supervised \ac{DL} has been widely considered to solve problems whose conventional solutions become too computationally complex, or the search spaces are too vast for considering approaches like genetic algorithms \cite{liu_deep_2020, elbir_deep_2020, yin_intelligent_2021, xia_deep_2020, ge_beamforming_2021}. However, the performance of supervised \ac{DL} is highly dependent on the data sets used for training, and the predictions are very sensitive to modifications in the communication systems \cite{goodfellow_deep_2016}.

Due to the limitations of the solutions above, recent works suggest the use of \ac{DRL}. Unlike supervised \ac{DL}, \ac{DRL} performs the training by interacting with the communication system. Hence, \ac{DRL} approaches do not require massive labeled sets of data. \ac{DRL}-based solutions have been employed in different scenarios for wireless communication systems, proving to be an appealing alternative when tackling complex optimization problems \cite{zhao_deep_2019,  huang_deep_2021, yang_intelligent_2021, nurani_krishnan_optimizing_2020, zhang_deep_2021}.
However, defining a \ac{MDP}\textemdash a requirement for \ac{RL} formulations\textemdash makes no sense for some optimization problems, and simpler formulations such as \ac{CB} could be more appropriate. 

According to \cite{sutton_reinforcement_2018}, \ac{CB} problems are intermediate between the $k$-armed bandit and \ac{RL} problems. \ac{CB} involves learning optimal actions but aiming to maximize an immediate reward instead of a long-term one. A significant difference with \ac{RL} is that actions and states in \ac{CB} are formulated so that actions do not affect the following states or rewards.

Recently, \ac{CB} has been gaining attention, and several \ac{CB}-based solutions have been proposed for different wireless communication open problems. In \cite{ismath_deep_2021}, the authors propose a \ac{DCB}-based instantaneous beam selection method for \ac{mmWave} cell-free networks. The results show that the proposed method outperforms other state-of-the-art algorithms in terms of latency and computational complexity, thus enabling faster initial access to the network. Authors in \cite{zhang_beam_2021} proposed an algorithm based on \ac{CB} for beam width selection in \ac{mmWave} links. Their proposal efficiently handled the fast channel variations by considering a codebook-aided approach. Authors in \cite{mohamed_sleeping_2021} also propose a \ac{CB}-based solution for \ac{mmWave} device-to-device two-hop relay probing, which outperforms the considered benchmarks in terms of relaying performance and execution time. 

\ac{CB}-based algorithms are an appealing alternative to handle the challenging scenarios of \ac{B5G} and \ac{6G}. However, as explained in \cite{Pan_2019}, standard \ac{CB} settings describe problems where optimal actions are selected from discrete sets of feasible actions. Hence, most existing works consider only discrete action spaces, which are unsuitable for joint optimization of the \ac{IRS} phase-shift matrix and the \ac{MIMO} precoders. This optimization problem becomes a combinatorial search problem with an intractable number of possible actions when considering a discrete set of actions. Additionally, unlike standard \ac{CB} settings, the actions in this optimization problem are expected to be composed of several configurable elements that must be jointly optimized. Therefore, the action space in this problem is high-dimensional, in the sense that many individual elements must be optimized, and continuous because each element is continuous-valued. These issues and the limitations of the scarce solutions in the literature have motivated our work, whose main contributions are:

\begin{itemize}
\item{The proposal of a new approach for the joint optimization of the IRS phase-shift and \ac{MIMO} precoder matrices, considering a CB formulation of the problem with continuous action and state spaces.}

\item{The development of an actor-critic framework called \ac{DCB-DDPG}, which enables handling the high-dimensional continuous action space by considering a deterministic policy gradient approach.}

\item {The use of a multi-head structure for the actor \ac{ANN} in the \ac{DCB-DDPG} framework. This way, the precoders' normalization, and the \ac{IRS} phase-shift matrix projection stages are performed within the actor. This proposal remarkably improves the training performance and is a major step for multi-agent implementations.}

\item {The design of a \ac{DRL}-based \ac{DDPG} formulation of the joint optimization problem where state and action spaces are selected in a way that is more appropriate to fit the requirements of the \ac{MDP}.}

\item{The evaluation of the performance of the proposed methods in an \ac{IRS}-assisted \ac{MS} \ac{MU-MIMO} system. We particularly focus on strong interference communication scenarios where the number of transmitted streams is larger than the number of receiving antennas, and the receiver does not have enough degrees of freedom to manage the multi-user interference. }

\end{itemize}

Hence, we propose an innovative approach to a well-known \ac{DRL}-based formulation (DRL-DDPG) and a game-changing approach based on DCB (DCB-DDPG). With a detailed analysis of both frameworks, we intend to understand better their strengths and weaknesses beyond those that can be inferred from observing the simulation results. The remainder of this paper is structured as follows. Section II details the theoretical fundamentals of the \ac{CB} and \ac{RL} frameworks and briefly analyzes some related works. Section III introduces the \ac{IRS}-assisted \ac{MS} \ac{MU-MIMO} system model and formulates the optimization problem to determine the \ac{IRS} phase-shift matrix and MIMO precoders. Sections IV and V explain the \ac{CB} and \ac{RL} proposed solutions, respectively. In Section VI, we present an analysis of their convergence and computational complexity. Section VII is devoted to the simulation results, and Section VIII presents the conclusions.

\section{Preliminaries}

\subsection{Reinforcement Learning (RL) and Contextual Bandit (CB)}
As stated in \cite{sutton_reinforcement_2018}, \ac{RL} is a computational approach to learning-by-interacting, i.e., mapping situations to actions that maximize a numerical reward function. Most \ac{RL} problems are formalized in terms of \acp{MDP}, where the agent (the learning and decision-maker element) interacts with the environment (all the elements external to the agent) through actions that affect the following states and rewards. Hence, \ac{RL} problems are described by a dynamics function $p(s_{t+1}, r_t | a_t, s_t)$ such that in every time instant $t$, the next state $s_{t+1}$, and the reward $r_t$ are conditioned by the effect of taking an action $a_t$ in the current state $s_t$. 

On the other hand, \ac{CB} problems can be understood as a relaxation of \ac{RL} problems. In \ac{CB}, interactions fit a dynamics function $p(r_t | a_t, s_t)$. Thus, only immediate rewards are affected by the current state and action. Unlike \ac{RL}, a state in \ac{CB} only contains the necessary information to select the best action for that state.

\begin{figure}[!t]
\centering
\includegraphics[width=0.8\columnwidth]{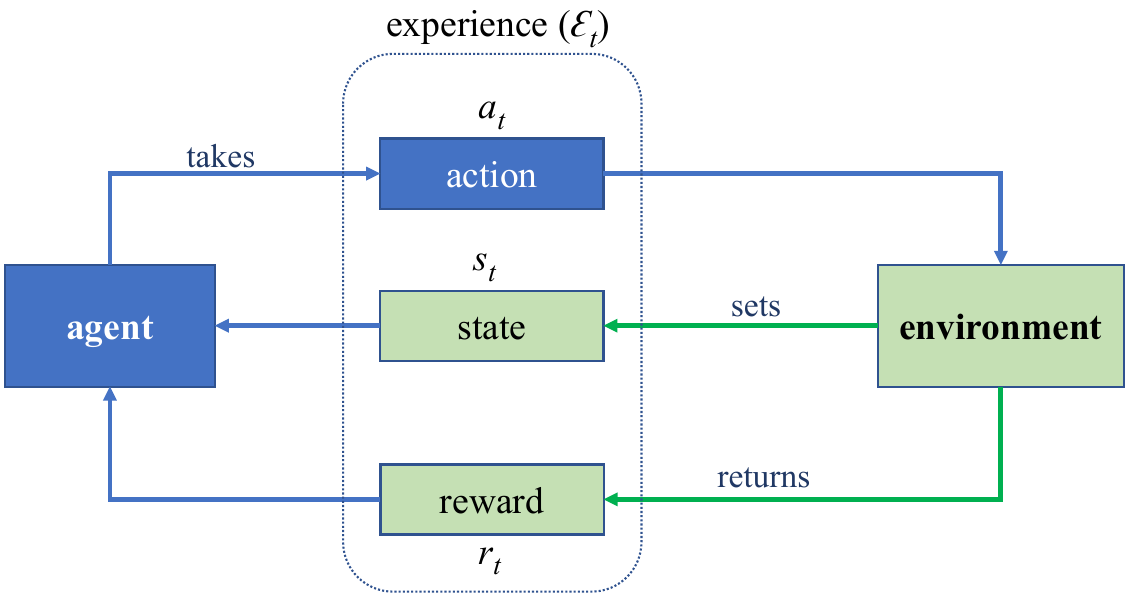}
\caption{Contextual bandit (CB) framework.}
\label{CBfig}
\end{figure}

\begin{figure}[!t]
\centering
\includegraphics[width=0.8\columnwidth]{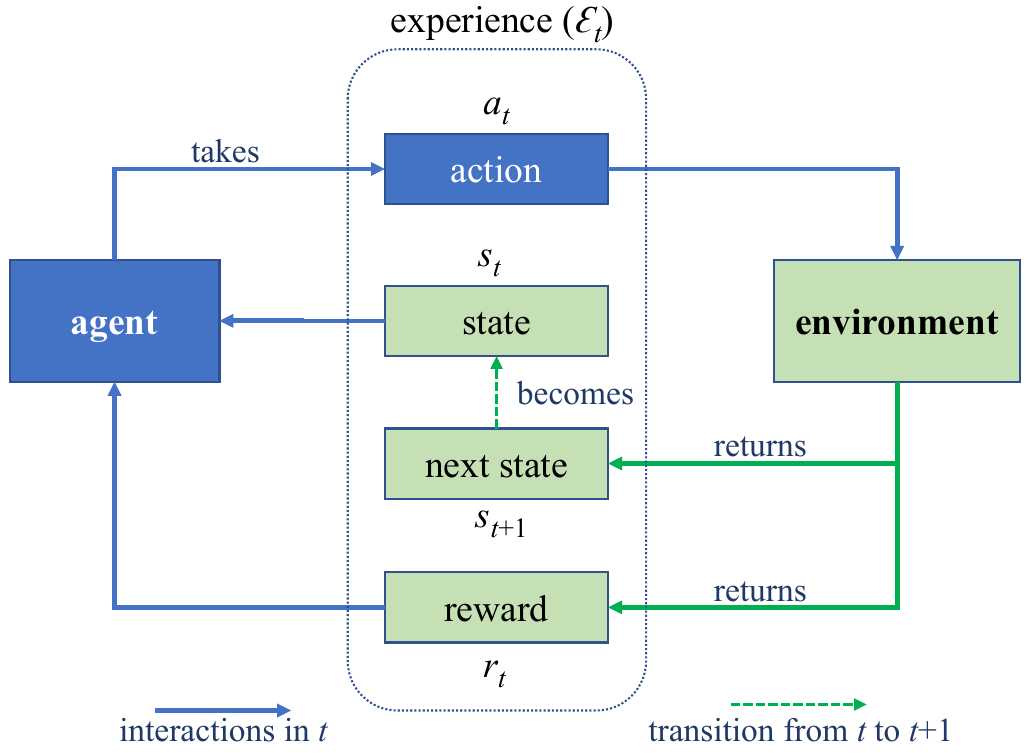}
\caption{Reinforcement learning (RL) framework.}
\label{RLfig}
\end{figure}

\Cref{CBfig} and \Cref{RLfig} show the interactions between the main components of the \ac{CB} and \ac{RL} frameworks, respectively. The environment is the external element with which the agents interact and sets the state $s_t$ in every instant $t$. However, in the \ac{CB} formulation, the selection of the current state is not related to the previous state and action. In both approaches, the agent selects the action $a_t$ that maximizes the value of a reward function, i.e., the immediate reward $r_t(\cdot)$ in \ac{CB} and a function of the expected long-term reward $q_\pi(\cdot)$ in \ac{RL}. The tuples $\mathcal{E}_t = (s_t, a_t, r_t)$ in \ac{CB} and $\mathcal{E}_t = (s_t, a_t, r_t, s_{t+1})$ in \ac{RL} are commonly termed experiences.

The policy is another critical element of both frameworks. It is the decision-making rule that defines the action that the agent will take while being in a given state, i.e., $a_t$ = $\pi$($s_t$), where $a_t$ belongs to the set $\mathcal{A} (s_t)$ of available actions in $s_t$, and $\pi (\cdot)$ is a deterministic policy. Hence, \ac{CB} and \ac{RL} algorithms aim to find the optimal policy $\pi^*(\cdot)$ that maximizes the respective reward functions. 

Conventional tabular approaches to \ac{CB} and \ac{RL} problems have proven efficient when considering low-dimensional, discrete state and action spaces \cite{sutton_reinforcement_2018}. However, these algorithms rapidly become intractable in problems where action and state spaces are continuous or arbitrarily large. In this regard, the \ac{ANN}-based \ac{DCB} and \ac{DRL} algorithms are appealing alternatives. The use of \acp{ANN} for function approximation of the policy and reward functions enables a wide range of new approaches to high-complexity problems like the one we are addressing in this work.

\subsection{Related Works}
In this subsection, we analyze two existing approaches to the joint optimization of the \ac{IRS} and precoder matrices. We focus on \cite{huang_reconfigurable_2020} and \cite{stylianopoulos_deep_2022}, which formulate the joint optimization as an \ac{RL} problem and a \ac{CB} problem, respectively. \Cref{tableRelW} summarizes some differences and similarities between these two approaches.

\begin{table*}[ht]
\centering
\caption{Related Works' Summary\label{tableRelW}}
\begin{tabular}{|c|c|c|ll}
\cline{1-3}
\multicolumn{1}{|c|}{Reference} & \cite{huang_reconfigurable_2020} & \cite{stylianopoulos_deep_2022}  &  &  \\ 
\cline{1-3}
Problem formulation & Reinforcement learning & Contextual bandit &  &  \\ 
\cline{1-3} State space  & \begin{tabular}[c]{@{}c@{}} Previous time step IRS matrix values,  previous time \\  step precoder values, CSI and power-related values\\   (Continuous-valued)\end{tabular} & \begin{tabular}[c]{@{}c@{}}CSI\\ (Continuous-valued)\end{tabular}  &  &  \\ 
\cline{1-3} Action space & \begin{tabular}[c]{@{}c@{}}Updated IRS matrix and precoder \\ (Continuous-valued)\end{tabular}                                  & \begin{tabular}[c]{@{}c@{}} Updated IRS matrices and precoder \\ (Discrete-valued)\end{tabular} &  &  \\ \cline{1-3}
\end{tabular}
\end{table*}

Authors in \cite{huang_reconfigurable_2020} propose an \ac{RL}-based \ac{DDPG} framework for the joint optimization of the IRS matrix and the precoder in the downlink of a single-stream \ac{MU-MISO} system. In their proposal, the action vectors are composed of the entries of the \ac{IRS} and precoder matrices. These matrices are considered continuous-valued, enabling high flexibility in searching for the optimal solution. Besides, it allows authors to overcome the high dimensionality of discrete-valued implementations. The instantaneous sum-rate value is the reward function, matching the metric considered for the optimization problem.

As shown in \Cref{tableRelW}, the received and transmitted powers are included in the states \cite{huang_reconfigurable_2020}. However, states also include the previous entries of the \ac{IRS} phase-shift and precoder matrices. Since the reward is a function of the current sum-rate value, it is unclear how the selection of the optimal matrices for a given channel realization could be related to the previously employed matrices. In \ac{RL}, the agent takes the action according to the state information. Hence, for a given channel realization, there would be as many possible optimal matrices as possible actions in the previous state. It seems more accurate to consider actions that only depend on the current channel realization and therefore rewards that only depend on the current channel realization and the \ac{IRS} and precoding matrices.

On the other hand, \cite{stylianopoulos_deep_2022} considers a \ac{CB}-based approach to the joint optimization of the precoders and the \ac{IRS} matrices in the downlink of a multi-\ac{IRS} MU-MISO system. As in \cite{huang_reconfigurable_2020}, the instantaneous sum-rate value is the reward. However, in \cite{stylianopoulos_deep_2022}, the state vectors are only composed of the channel coefficients for the current channel realization, which properly follows the \ac{CB} formulation.

The major drawback of the solution proposed in \cite{stylianopoulos_deep_2022} is the assumption of only discrete-valued actions, i.e., IRS scattering elements with only two possible phase values and a codebook set for the precoder selection. In setups like that, with low-dimensional sets of possible actions, the system performance is limited, and the benefits of \ac{IRS}-assisted \ac{MIMO} communications cannot be fully leveraged. On the other hand, the complexity of the proposed algorithm steeply increases with the number of possible actions. This issue is interpreted in \cite{lillicrap_continuous_2019, sutton_reinforcement_2018} as an instance of the curse of dimensionality.

\subsection{Overcoming the Limitations}

Based on the previous analysis of related works, we propose two different approaches to overcome their limitations. The first is a \ac{CB} formulation of the joint optimization problem where action and state spaces are considered to be continuous. To handle this situation, we introduce a novel framework called \ac{DCB-DDPG}. It leverages several features of \ac{DDPG} and adapts them to the \ac{CB} formulation. As introduced in \cite{lillicrap_continuous_2019}, \ac{DDPG} is a model-free, off-policy actor-critic framework based on the \ac{DPG} algorithm. \ac{DDPG} has rapidly become a well-established solution to high-dimensional, continuous action spaces and its features make it a suitable alternative to assist our \ac{DCB} formulation of the joint optimization problem \cite{huang_reconfigurable_2020, xu_experience-driven_2021,pereira-ruisanchez_joint_2022, albinsaid_multi-agent_2021, guo_learning-based_2021}.

In this sense, model-free algorithms enable learning from interactions in scenarios where system models become too complex or distant from the actual behavior of practical deployments. Although the initial training of the proposed solution is based on the outcomes of the IRS-assisted MS MU-MIMO uplink system model, its performance during practical evaluations will depend on its capability to keep learning from interacting. Besides, the off-policy behavior enables exploring more efficiently high-dimensional search spaces like ours and using past experiences during training. Off-policy algorithms can be trained with either externally generated or self-generated experiences, thus reducing the possibility of getting stuck in a local minimum \cite{sutton_reinforcement_2018}. Finally, the \ac{DPG} updates are suitable for our aim of finding a deterministic policy that returns the continuous-valued optimal action in every continuous-valued state while keeping the evaluation variance as low as possible \cite{silver_deterministic_2014}. 

From the above analysis, the second proposal is a \ac{DRL}-based \ac{DDPG} approach. However, we consider sets of states and actions that differ from those considered in the related work \cite{huang_reconfigurable_2020} and, as discussed in later sections, can be more appropriate to fit the requirements of the \ac{MDP}. For greater clarity, this second framework will be termed DRL-DDPG.

During the initial phases of the investigation, we analyzed other \ac{DRL}-based algorithms that lacked some of the desired features above, i.e., they were either on-policy methods or learned stochastic policies (e.g., \ac{PPO}, \ac{TRPO} and \ac{A3C} algorithms). In on-policy methods, exploration is restricted to the learned target policy plus a certain degree of randomness in the actions taken, which is a limiting factor in high-dimensional continuous-valued search spaces \cite{sutton_reinforcement_2018}. On the other hand, stochastic policies are more desirable in \ac{POMDP} formulations of \ac{RL} problems, where the environment uncertainty characterizes the interactions \cite{zai_deep_2020, sutton_reinforcement_2018}. However, in our formulation, the relationship between the observed states, actions, and rewards is established deterministically: system sum-rate values are unequivocally related to the channel realizations and the configured IRS matrix and precoders. In addition, to improve the performance of our deterministic approach, we consider several mechanisms that enable efficient exploration of both the action and the state spaces.

For a practical evaluation of both proposed approaches, the considered system model has been extended to \ac{MS} \ac{MU-MIMO} communications. Additionally, we address several scenarios where there are not enough degrees of freedom to cancel all the multi-user interference. 

\hspace{5mm}
\section{System Model and Optimization Problem}

\subsection{Notation}
Along this work, the following notation will be employed: $a$ is a scalar, $\mathbf{a}$ is a vector, and $\mathbf{A}$ represents a matrix. Transpose, conjugate transpose, and the Frobenius norm of $\mathbf{A}$ are represented by $\mathbf{A}^\text{T}$, $\mathbf{A}^\text{H}$, and ${\|\mathbf{A} \|}_\text{F}^2$, respectively. Calligraphic letters are employed to denote sets and tuples. $|\mathcal{R}|$ stands for the cardinality of a set $\mathcal{R}$. $\mathbf{I}_N$ indicates an $N \times N$ identity matrix, and $\mathcal{I}_N$ denotes the set of integers from 1 to $N$. The operator $\operatorname{blkdiag\;(\cdot)}$ constructs a block diagonal matrix from its input matrices. Finally, $\operatorname{flatten}(\cdot)$ is the operator that reshapes any matrix $\mathbf{V}\in \mathbb{C}^{A\times B}$ into a vector $\mathbf{v}\in \mathbb{C}^{1\times AB}$ by concatenating all the entries. The mathematical relationships presented in the following sections hold for all the consecutive time steps $t$ that fit within one coherence block. Hence, for the sake of simplicity, sub-index $t$ is used only where necessary to avoid ambiguities. 

\subsection{IRS-assisted MS MU-MIMO Uplink}

\begin{figure}[!t]
	\centering
	\includegraphics[width=1\columnwidth]{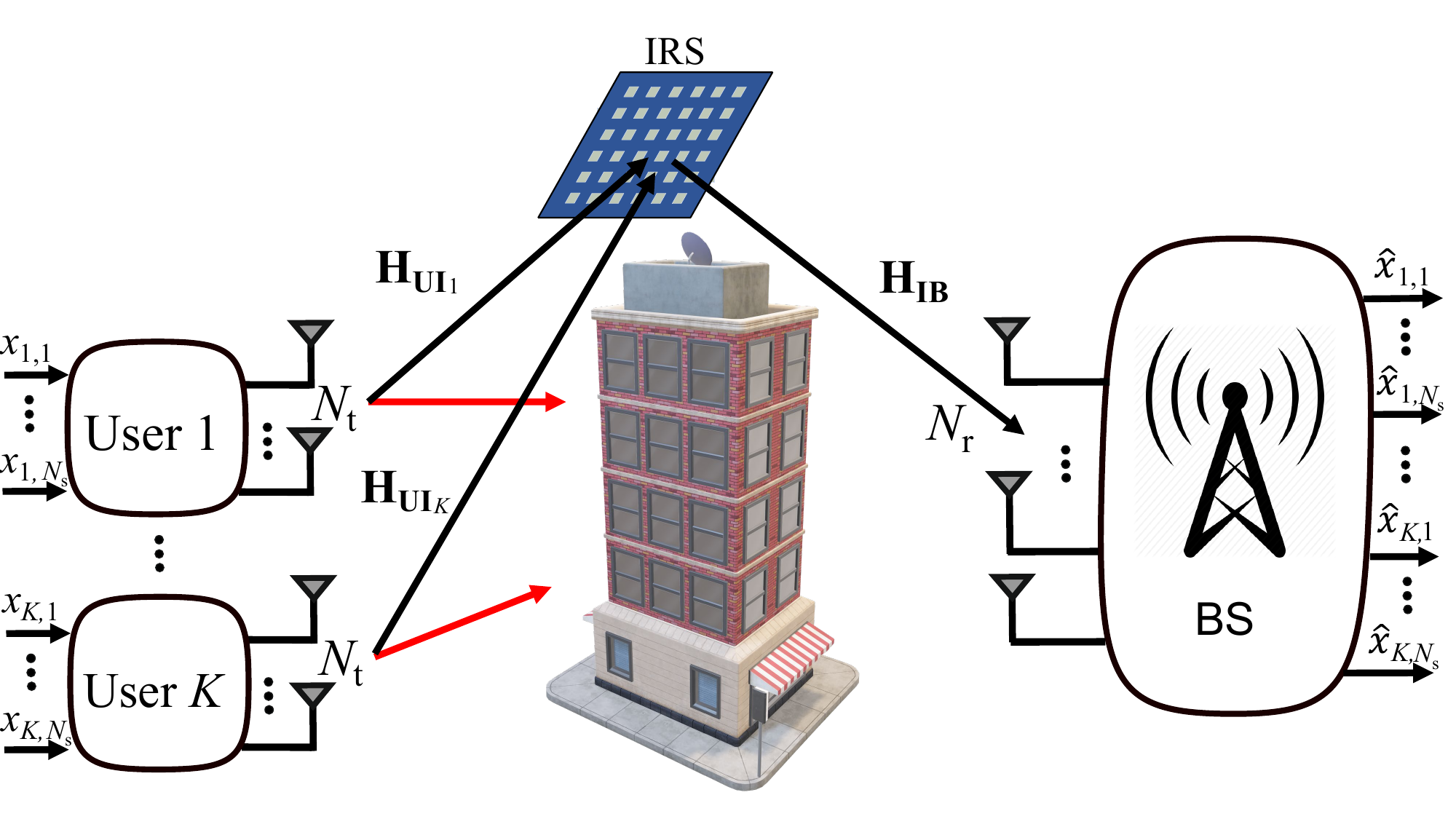}
	\caption{Uplink of an \ac{IRS}-assisted  \ac{MS} \ac{MU-MIMO} system.}
	\label{SystemModel}
\end{figure}

Let us consider the uplink of an \ac{IRS}-assisted \ac{MS} \ac{MU-MIMO} communication system, as shown in \Cref{SystemModel}. In this scenario, each of the $K$ users employs $N_{\text{t}}$ antennas to send $N_{\text{s}}$ data streams to a \ac{BS} equipped with $N_{\text{r}}$ antennas. We assume there is a total blockage between the users and the \ac{BS}, i.e., no direct channel is available between any of the $K$ users and the \ac{BS}.     

The symbols transmitted by the $k$-th user on each channel usage are represented by $\mathbf{x}_k = [x_{k,\text{1}}, \ldots, x_{k,N_{\text{s}}}]^\text{T}\in \mathbb{C}^{N_{\text{s}}\times 1 }$. We assume $\mathbf{x}_k$ follows a zero-mean multivariate complex-valued Gaussian distribution, i.e., $\B{x}_k\sim \mathcal{N}_{\mathbb{C}} (0, \B{I}_{N_\text{s}})$. Such symbols are linearly processed with the precoder ${\mathbf{P}_{}}_{k} \in \mathbb{C}^{N_{\text{t}}\times N_{\text{s}}}$. The power transmitted by the $k$-th user is limited to $\Omega_k$, which leads to the individual power constraint on the precoders given by $ \Vert  {{\mathbf{P}}_{}}_{k} \Vert_\text{F}^2 \leq \Omega_k,\;\forall k$.

The deployed \ac{IRS} is assumed to have $N$ scattering elements. Thus, the \ac{IRS} phase-shift matrix is represented by the diagonal matrix $\mathbf{\Theta}=\text{diag}(e^{j\theta_1},\ldots,e^{j\theta_{N}})\in\mathcal{D}$ where $\theta_{n}\in[0,2\pi)$ represents the phase shift introduced by the $n$-th element of the \ac{IRS}. $\mathcal{D}\in \mathbb{C}^{N\times N} $ is the set of feasible \ac{IRS} matrices, i.e., the diagonal matrices with unit modulus entries. The vector that contains the entries in the main diagonal of $\mathbf{\Theta}$ is denoted by $\boldsymbol{\theta}\in \mathbb{C}^{N\times 1}$.  

According to this system model, the received signal at the \ac{BS} is given by 
\begin{equation}\label{1}
\mathbf{y}={\mathbf{{H}}_{\text{IB}}}_{}\mathbf{\Theta}\sum_{k=1}^{K}{{\mathbf{{H}}_{\text{UI}}}_{k}} {\mathbf{P}_{}}_{k}  \mathbf{x}_{k}+\mathbf{{n}},
\end{equation}
where the matrix ${\mathbf{{H}}_{\text{UI}}}_{k}\in{\mathbb{C}^{N_{\text{}}\times N_{\text{t}}}}$ represents the channel response of the link from the $k$-th user to the \ac{IRS}, ${\mathbf{{H}}_{\text{IB}}}\in{\mathbb{C}^{N_{\text{r}}\times N_{\text{}}}}$ stands for the channel response from the \ac{IRS} to the \ac{BS}, and $\B{{n}}=\left[n_{1}, n_{2},\ldots,n_{N_{\text{r}}}\right]^\text{T}$ represents the complex-valued \ac{AWGN} which is modeled as $\B{n}\sim \mathcal{N}_{\mathbb{C}} (0, \sigma_{\text{n}}^2\B{I}_{N_{\text{r}}})$.

Using a more compact notation, the received signal at the \ac{BS} given by \eqref{1} can be rewritten as
\begin{align}
\mathbf{y}={\mathbf{{H}}_{\text{IB}}}_{}\mathbf{\Theta}\mathbf{H}_{\text{UI}}\mathbf{P}\mathbf{x}+\mathbf{n},
\end{align} 
where $\mathbf{H}_{\text{UI}}=[{\mathbf{{H}}_{\text{UI}}}_{1},\ldots,{\mathbf{{H}}_{\text{UI}}}_{K}]\in \mathbb{C}^{{N}\times KN_{\text{t}}}$, $\mathbf{x} = [\mathbf{x}_{\text{1}}^\text{T}, \ldots, \mathbf{x}_{K}^\text{T}]^\text{T}\in \mathbb{C}^{KN_{\text{s}}\times 1}$ and $\mathbf{P}=\operatorname{blkdiag}({\mathbf{P}_{\text{}}}_1,\ldots,{\mathbf{P}_{\text{}}}_K)\in \mathcal{P}$ is the block-diagonal matrix which stacks all the users' precoders. $\mathcal{P}\in\mathbb{C}^{N_{\text{t}}K\times N_{\text{s}}K}$ is the set of block-diagonal matrices with norm-constrained submatrices.

The vector with all the estimated user symbols $\hat{\mathbf{x}} = [\hat{\mathbf{x}}_{\text{1}}^\text{T}, \ldots, \hat{\mathbf{x}}_{K}^\text{T}]^\text{T}\in \mathbb{C}^{KN_{\text{s}}\times 1}$ is obtained by the linear filtering of the signal received at the \ac{BS}, i.e., $\hat{\mathbf{x}}=\mathbf{W}^\text{H}\mathbf{y}$, where $\mathbf{W}^\text{H}=[\mathbf{W}_{1},\ldots,\mathbf{W}_{K}]^\text{H}\in \mathbb{C}^{KN_{\text{s}}\times N_{\text{r}}}$ is the \ac{BS} receiving filter matrix that stacks the individual receiving filter matrices $\mathbf{W}^\text{H}_{k}\in \mathbb{C}^{N_{\text{s}}\times N_{\text{r}}}$. We assume ${\mathbf{W}}^\text{H}_{k},\;\forall k$ to be the \ac{MMSE} receiving filter given by
\begin{align}\label{MMSE}
&\mathbf{W}^\text{H}_{k}=\mathbf{P}^\text{H}_k{\mathbf{H}^\text{H}_{\text{UI}}}_k\mathbf{\Theta}^\text{H}\mathbf{{H}}_{\text{IB}}^\text{H}
\\&\times{\left({\mathbf{{H}}_{\text{IB}}}_{}\mathbf{\Theta}\mathbf{H}_{\text{UI}}\mathbf{P}\mathbf{P}^\text{H}\mathbf{H}_{\text{UI}}^\text{H}\mathbf{\Theta}^\text{H}\mathbf{{H}}_{\text{IB}}^\text{H}+\sigma^2_{\text{n}}\mathbf{I}_{N_{\text{r}}}\right)}^{-1}.
\notag
\end{align}

Considering this system model, we look for the \ac{IRS} phase-shift matrix and the precoders that maximize the achievable sum-rate. Towards this aim, we formulate the following optimization problem
\begin{align}\label{OpA}
&\underset {\mathbf{P}, \mathbf{\Theta}} {\arg \max} \sum_{k=1}^{K}~{R_{k}}\\
\text{s.t.}\;&{\parallel {\mathbf{P}_{}}_k} \parallel_\text{F}^2\leq \Omega_{k},\;\forall k\in\mathcal{I}_K, \notag\\  & \mathbf{\Theta}\in\mathcal{D},\notag
\end{align}
where $R_k$ is the $k$-th user individual rate which is given by
\begin{align}\label{Sr}
&{R_{k}} = {{\text{log}_{2}\;\text{det}\;\big(\mathbf{I}_{{K}}+\mathbf{X}_k^{-1}}}\\
&\times \mathbf{W}^\text{H}_{k}\;
{\mathbf{H}^{}_{\text{IB}}}\mathbf{\Theta}{\mathbf{H}^{}_{\text{UI}}}_k{\mathbf{P}^{}}_{k}{\mathbf{P}}_{k}^\text{H}{\mathbf{H}^\text{H}_{\text{UI}}}_k\mathbf{\Theta}^\text{H}{\mathbf{H}^\text{H}_{\text{IB}}}\mathbf{W}_{k}\big),\notag
\end{align}
where
\begin{align}
&\mathbf{X}_k\hspace{-1mm}=\sum_{i\neq k}{ \mathbf{W}^\text{H}_{k}
{\mathbf{H}_{\text{IB}}}\mathbf{\Theta}{\mathbf{H}_{\text{UI}}}_i{{\mathbf{P}^{}}_{i}{\mathbf{P}}_{i}^\text{H}}{\mathbf{H}^\text{H}_{\text{UI}}}_i\mathbf{\Theta}^\text{H}{\mathbf{H}^\text{H}_{\text{IB}}}\mathbf{W}_{k}}\\
&
+\sigma^2_{\text{n}}{\mathbf{W}^\text{H}_{k}}{\mathbf{W}_{k}}\notag
\end{align}
is the interference plus noise matrix, and ${\mathbf{W}}^\text{H}_k$ is the \ac{MMSE} individual receiving filter for the $k$-th user, as in \eqref{MMSE}. Notice that for the single-stream scenarios, $N_\text{s}=1$, the user individual receiving filter matrices ${\mathbf{W}}^\text{H}_k\in\mathbb{C}^{N_{\text{s}}\times N_{\text{r}}}$ collapse in the vectors ${\mathbf{w}}^\text{H}_k\in\mathbb{C}^{1\times N_{\text{r}}}$.
 
As stated in \eqref{OpA}, the precoders must meet the individual power constraints ${\parallel {\mathbf{P}_{}}_k} \parallel_\text{F}^2\leq \Omega_{k},\;\forall k$, where $\Omega_k$ represents the available power at the k-th user. For the sake of simplicity, we assume the same power constraint value for all the users, i.e., $\Omega_k=\Omega,\;\forall k$. Without loss of generality, we also assume a noise variance $\sigma^2_{\text{n}}$ equal to one. Therefore, the \ac{SNR} per user is given by $\text{SNR}_\text{(dB)}=10 \;\text{log}_{10} (\Omega)$.

\begin{table}[!t]
\caption{System Model Parameters\label{tableSystModel}}
\centering
\begin{tabular}{|c|c|}
\hline
Parameter &Description\\
\hline
$K$ & number of users\\
\hline
$N_\text{t}$ & number of transmitting antennas per user\\
\hline
$N_\text{s}$ & number of transmitted streams per user\\
\hline
$N_\text{r}$ & number of receiving antennas at BS\\
\hline
$\mathbf{H}_{\text{UI}}$ & channel response matrix from the users to the IRS\\
\hline
$\mathbf{H}_\text{IB}$ & channel response matrix from the IRS to the BS\\
\hline
$\mathbf{P}_{k}$ & precoder matrix for each user $k$\\
\hline
$\mathbf{P}$ & block-diagonal matrix with all the users' precoders\\
\hline
$\mathbf{\Theta}$ & IRS phase-shift matrix\\
\hline
$\boldsymbol{\theta}$ & vector of the IRS phase-shifts\\
\hline
$\Omega_{k}$ & users' individual power constraint\\
\hline
${\mathbf{W}}^\text{H}_{k}$ & MMSE receiving filter for each user $k$\\
\hline
${\mathbf{X}}_{k}$ & interference plus noise matrix for each user $k$\\
\hline
$R_k$ & achievable rate for user $k$\\
\hline
\end{tabular}
\end{table}

\Cref{tableSystModel} summarizes the main system model parameters and their descriptions. Due to the choice of the objective function and the design constraints for the \ac{IRS} and the users' precoders, \eqref{OpA} becomes a non-convex, non-trivial optimization problem. We propose using the DCB-DDPG and DRL-DDPG frameworks to handle this problem efficiently. Although we will only address uplink communication scenarios like the one previously described, proposed \ac{CB} and \ac{RL} solutions can be easily adapted to downlink scenarios, and similar results are obtained.

\subsection{Channel Model}
Channel responses from the users to the \ac{IRS} ($\mathbf{H}_{\text{UI}}=[{\mathbf{H}_{\text{UI}}}_{1},\ldots,{\mathbf{{H}}_{\text{UI}}}_{K}]$) and from the \ac{IRS} to the \ac{BS} (${\mathbf{{H}}_{\text{IB}}}$) are assumed to be perfectly known, as in \cite{wu_beamforming_2020, ismath_deep_2021, stylianopoulos_deep_2022, huang_deep_2021, yang_intelligent_2021}. Although channel acquisition is challenging in IRS-aided systems, a promising line of research is committed to facing this problem \cite{wang_channel_2020, joham_estimation_2022, you_channel_2020}. Therefore, it is reasonable to assume that the \ac{IRS}-assisted communication systems can implement some appropriate mechanism to provide accurate \ac{CSI} to the agents in the \ac{DCB} and \ac{RL} frameworks.

In particular, we assume a Rayleigh fading model for ${\mathbf{{H}}_{\text{UI}}}_{k}$, i.e., the entries of ${\mathbf{{H}}_{\text{UI}}}_{k},\;\forall k$ are independent and identically distributed (i.i.d.) complex-valued circularly symmetric Gaussian random variables. On the other hand, considering that \acp{IRS} are typically installed to ensure \ac{LoS} to the \ac{BS}, a Rician fading channel model is adopted to describe ${\mathbf{{H}}_{\text{IB}}}$ as in \cite{wu_beamforming_2020, stylianopoulos_deep_2022}, i.e.,
\begin{equation}
    \mathbf{H}_{\text{IB}}=\sqrt{\frac{\beta}{1+\beta}}\mathbf{H}_{\text{IB}}^{\text{LOS}}+\sqrt{\frac{1}{1+\beta}}\mathbf{H}_{\text{IB}}^{\text{NLOS}},
\end{equation}
where $\beta$ is the Rician factor, which  is  set  to  3  dB. $\mathbf{H}_{\text{IB}}^{\text{LOS}}$ and $\mathbf{H}_{\text{IB}}^{\text{NLOS}}$ stand for the \ac{LoS} component and the Rayleigh fading component, respectively. 

\section{CB-based Joint Optimization}
In this section, the optimization problem in \eqref{OpA} is solved using a novel \ac{CB}-based approach. We start selecting the state, action, and reward spaces. Next, we review some essential components in the proposed \ac{DCB-DDPG} framework, and finally, we derive the algorithmic solution.

\subsection{DCB-DDPG: State, Action and Reward}
According to the \ac{CB} formulation, we introduce the following states, actions, and rewards:
\begin{itemize}
\item{The \textbf{state} vector $\mathbf{s}_{t}$ comprises the current values of all the channel response matrices from the users to the \ac{IRS} (${\mathbf{{H}}_{\text{UI}}}_{k},\;\forall k$) and from the \ac{IRS} to the \ac{BS} (${\mathbf{{H}}_{\text{IB}}}$). The state vector is constructed such that 
\begin{align}\label{stateDCB}
\mathbf{s}_{t}=[\operatorname{flatten}({\mathbf{{H}}_{\text{UI}}}_{1}),\ldots,\operatorname{flatten}({\mathbf{{H}}_{\text{UI}}}_{K}),\\\operatorname{flatten}({\mathbf{{H}}_{\text{IB}}})].\notag
\end{align}
Hence, the dimension of the state space vectors is $D_\text{state}^{\text{DCB}}= KN_\text{t}N+N_\text{r}N$. Recall that the state space is continuous since the entries of $\mathbf{s}_{t}$ can take any complex value. Notice that we assume signal processing techniques that enable handling complex-valued entries. Otherwise, the imaginary and real parts must be treated as independent inputs, leading to vectors twice the size.}  
\item{The \textbf{action} vector $\mathbf{a}_{t}$ is composed of the entries in the main diagonal of the \ac{IRS} phase-shift matrix ($\boldsymbol{\theta}$) and those in all the users' precoders ($\mathbf{P}_k,\;\forall k$). Hence, the action vector is constructed such that 
\begin{align}\label{actionDCB}
\mathbf{a}_{t}=[\operatorname{flatten}(\mathbf{P}_1),\ldots,\operatorname{flatten}(\mathbf{P}_K),\\\operatorname{flatten}(\boldsymbol{\theta})].\notag
\end{align}

The dimension of the action space vectors is $D_\text{action}^{\text{DCB}} = KN_\text{t}N_\text{s} + N$. The action space is also continuous, though the matrices formed by these entries must meet the previously defined constraints (${\parallel {\mathbf{P}_{}}_k} \parallel_\text{F}^2\leq \Omega,\;\forall k,$ and $\mathbf{\Theta}\in\mathcal{D}$).}

\item{The \textbf{reward} $r_{t}$ is equal to the system sum-rate determined according to \eqref{OpA}, and by considering the entries of the current state and the action vectors, i.e.,  
\begin{align}\label{rewardDCB}
{r}_{t}=\sum_{k=1}^{K}~{R_{k}}(\mathbf{s}_{t}, \mathbf{a}_{t}).
\end{align}}
\end{itemize}

\subsection{DCB-DDPG: Framework Elements}
\begin{figure}[!t]
	\centering
	\includegraphics[width=0.9\columnwidth]{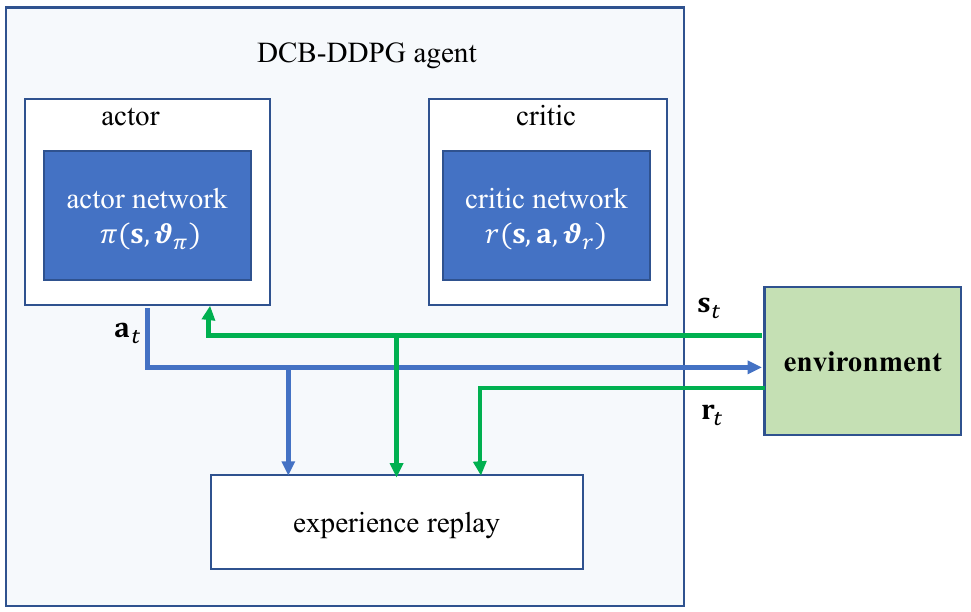}
	\caption{Elements in the \ac{DCB-DDPG} agent.}
	\label{DCB_DDPGframework}
\end{figure}

\ac{CB} algorithms are mostly oriented to solve discrete action space formulations, such as those presented in \cite{stylianopoulos_deep_2022, ismath_deep_2021, zhang_beam_2021, mohamed_sleeping_2021}. Hence, to address the high-dimensional continuous action space of the \ac{CB}-based formulation presented in the previous subsection, we have developed the proposed \ac{DCB-DDPG} framework. \Cref{DCB_DDPGframework} shows a representation of the agent that we consider in our proposal. Some of the elements that support this setup are the following: 

\begin{itemize}
\item{Actor-critic: called this way due to the interactions between the reward and policy functions. The critic is the element that updates the reward function according to the observed instantaneous rewards. Meanwhile, the actor updates the policy function by learning the actions that maximize the reward function. The update of the actor function is made through deterministic policy gradient updates as described in \cite{silver_deterministic_2014, lillicrap_continuous_2019}. Iteratively, both converge to the corresponding optimal function.}

\item{\acp{ANN} for function approximation: the reward and the policy functions are represented through \acp{ANN}. This way, the continuous action and state spaces can be efficiently handled. The actor network $\pi(\mathbf{s},\boldsymbol{\vartheta}_\pi)$ is the approximation of the policy function, and $\boldsymbol{\vartheta}_\pi$ is the vector of weights of the actor network to be trained. Similarly, $r(\mathbf{s}, \mathbf{a},\boldsymbol{\vartheta}_\text{r})$ is the approximation of the reward function, and $\boldsymbol{\vartheta}_\text{r}$ is the vector of weights of the critic network.}

\item{Experience replay: the experience tuples $\mathcal{E}_t$ observed at each time step are stored in a replay memory $\mathcal{R}$, such that they can be used for off-policy training. $|\mathcal{R}|$ is the number of experience tuples stored, and random mini-batches $\mathcal{B}$ with size $|\mathcal{B}|$ are used to update the weights of the actor and critic networks during training. Using the experience replay improves the stability of the learning stages and eliminates the undesirable effects of the correlation between samples.}
\end{itemize}

\subsection{DCB-DDPG: Proposed Algorithm}
In this section, we present the algorithm proposed to train the DCB-DDPG agent for solving the optimization problem in \eqref{OpA}. The values of the \ac{IRS} phase-shift matrix ($\mathbf{\Theta}$) and the precoder ($\mathbf{P}$) that maximize the achievable sum-rate are learned by following the steps in \Cref{DCB_alg}.

\begin{algorithm}[ht!]
\caption{DCB-DDPG algorithm}
\begin{algorithmic}[1]
\STATE \hspace{0.1cm}$ \textbf{Initialize: }$
\STATE \hspace{0.1cm}$ \text{set }\pi(\mathbf{s},\boldsymbol{\vartheta}_\pi) \text{ given random } \boldsymbol{\vartheta}_\pi$
\STATE \hspace{0.1cm}$\text{set }r(\mathbf{s}, \mathbf{a},\boldsymbol{\vartheta}_\text{r}) \text{ given random } \boldsymbol{\vartheta}_\text{r}$ 
\STATE \hspace{0.1cm}$ \text{create } \mathcal{R}$
\STATE \hspace{0.1cm}$ \textbf{for }t = 0,\ldots, T-1\textbf{ do:} $
\STATE \hspace{0.5cm}$ \text{set }\mathbf{s}\textsubscript{t} \text{ given channels }{\mathbf{{H}}_{\text{UI}}}_{k},\forall k\text{ and } {\mathbf{{H}}_{\text{IB}}}  $
\STATE \hspace{0.5cm}$ \text{agent takes }\mathbf{a}_{t} = \pi(\mathbf{s}_{t},\boldsymbol{\vartheta}_\pi) + \mathbf{n}_\text{e}$
\STATE \hspace{0.5cm}$ \text{environment returns }r_{t}$
\STATE \hspace{0.5cm}$ \mathcal{R} \text{ stores } \mathcal{E}_t = (\mathbf{s}_{t}, \mathbf{a}_{t}, r_{t})$
\STATE \hspace{0.5cm}$ \textbf{if }|\mathcal{R}| > |\mathcal{B}|:$
\STATE \hspace{1.0cm}$ \text{agent samples }|\mathcal{B}|\text{ random experiences: }$
\STATE \hspace{1.0cm}$ \mathcal{E}_i=(\mathbf{s}_{i}, \mathbf{a}_{i}, r_{i}), ~i=0,\ldots, |\mathcal{B}|-1 $
\STATE \hspace{1.0cm}$ \text{calculate the critic loss } L_{\text{c}} \text{ by using \eqref{LcDCB}} $
\STATE \hspace{1.0cm}$ \text{update } \boldsymbol{\vartheta}_\text{r} \text{ by the back-propagation of }L_{\text{c}}$
\STATE \hspace{1.0cm}$ \text{calculate the actor loss } L_{\text{a}} \text{ by using \eqref{LaDCB}} $
\STATE \hspace{1.0cm}$ \text{update } \boldsymbol{\vartheta}_\pi \text{ by the back-propagation of }L_{\text{a}}$
\STATE \hspace{0.5cm}$ \textbf{end if }$
\STATE \hspace{0.1cm}$ \text{Obtain } \mathbf{P} \text{ and } \mathbf{\Theta} \text{ by evaluating policy }\pi(\mathbf{s},\boldsymbol{\vartheta}_\pi)$
\end{algorithmic}
\label{DCB_alg}
\textbf{Output: } \text{trained actor network }$\pi(\mathbf{s},\boldsymbol{\vartheta}_\pi)$
\end{algorithm}

We first create the actor and critic networks by randomly initializing the network parameters ($\boldsymbol{\vartheta}_\pi$ and $\boldsymbol{\vartheta}_\text{r}$, respectively). We also create the actor network optimizer and the critic network optimizer to handle the updates of $\boldsymbol{\vartheta}_\pi$ and $\boldsymbol{\vartheta}_\text{r}$, respectively. The experience replay buffer $\mathcal{R}$ is created with no elements.

We propose to split the interactions between the \ac{DCB-DDPG} agent and the environment into $T$ time steps. Since no terminal states can be defined in our problem, the number of time steps can be arbitrarily selected. The state $\mathbf{s}_{t}$ is created as in \eqref{stateDCB} from randomly generated ${\mathbf{{H}}_{\text{UI}}}_{k},\;\forall k$ and ${\mathbf{{H}}_{\text{IB}}}$ channel response matrices. Hence, episodes are not required in this \ac{CB} formulation since the randomness of the channel response matrices guarantees continuous exploration over the state space.

From the current state $\mathbf{s}_{t}$, the agent selects the action $\mathbf{a}_{t}$ according to the output of the actor network and an exploration noise $\mathbf{n}_\text{e}\sim \mathcal{N}_{\mathbb{C}}(0, \sigma_{\text{n}_\text{e}}^2\B{I}_{D_\text{action}^{\text{DCB}}})$. Incorporating this noise improves the exploration of the action space since actions around the one selected by the actor network are evaluated. As explained in \cite{sutton_reinforcement_2018, lillicrap_continuous_2019, feng_deep_2020}, in off-policy algorithms, the policy resulting from adding the noise can be treated as a different policy. Samples for training can be obtained by following any policy, and the learning capability is not affected.

The learning stage (lines 11 to 16 in Algorithm \ref{DCB_alg}) is first performed when there are enough stored experiences to sample a mini-batch of size $|\mathcal{B}|$. Notice that the first mini-batches will have several samples in common since $|\mathcal{B}| \approx |\mathcal{R}|$. However, this effect soon disappears as $|\mathcal{R}|$ increases in every time step.

The experience tuples in $\mathcal{B}$ are employed to train the actor and critic networks as described in lines 13 to 16. This algorithm aims to maximize the instantaneous reward, not a long-term reward function, as in \ac{RL}. Hence, the sampled reward values are used as targets for the critic training. The critic loss, $L_{\text{c}}$, is then calculated as
\begin{align}\label{LcDCB}
L_{\text{c}} = \frac{1}{|\mathcal{B}|}\sum_{i} ({r}_{i}-r(\mathbf{s}_{i}, \mathbf{a}_{i},\boldsymbol{\vartheta}_\text{r}))^2
\end{align} 
and the obtained values are back-propagated by using the critic network optimizer. This way, the critic network is trained to minimize $L_{\text{c}}$ and represent the behavior of the actual reward function as closely as possible. 

Later, by following a deterministic policy gradient update, the actor network is trained to predict the action that maximizes the output of the critic network. Because of this, the sign of the critic output value is changed, i.e., 
\begin{align}\label{LaDCB}
L_{\text{a}} = -\frac{1}{|\mathcal{B}|}\sum_{i} r(\mathbf{s}_{i}, \pi(\mathbf{s}_{i},\boldsymbol{\vartheta}_\pi),\boldsymbol{\vartheta}_\text{r}),
\end{align} 
and, hence, the back-propagation of the actor network optimizer can be perceived as a gradient ascent instead of a descent. Notice that $L_{\text{a}}$ is intended only to update the actor network parameters. Thus, the actor network optimizer does not update the critic network parameters. 

At the end of every learning time step, the performance of the obtained policy is tested by running an evaluation step. Only unseen channel realizations are visited during the evaluation, and no learning or exploration is performed. The evaluation stage has two main aims: the detection of undesired behaviors during the learning stage, which could lead to performance degradation, and the identification of the policies that perform better with the unseen channel responses so that they can be used in the operational stage. The expected result is to obtain a trained policy network capable of predicting near-optimal \ac{IRS} and precoder matrices.

\subsection{DCB-DDPG: \acp{ANN} Structure} 

\begin{figure}[!t]
	\centering
	\includegraphics[width=0.9\columnwidth]{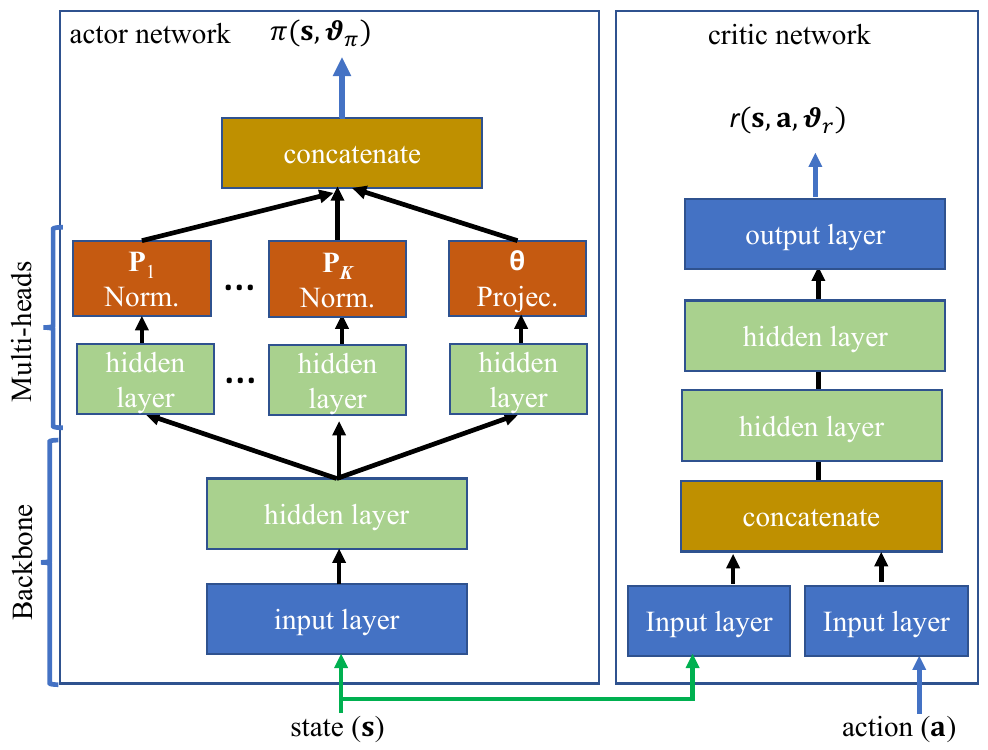}
	\caption{DCB-DDPG actor and critic network structure.}
	\label{DCB_ANN}
\end{figure}

\cref{DCB_ANN} shows the structures of the \acp{ANN} that we propose to use as the actor and critic networks. The actor network input and output layer dimensions equal $D_\text{state}^{\text{DCB}}$ and $D_\text{action}^{\text{DCB}}$, respectively. For this network, we employ a multi-head \ac{ANN} setup such that each user precoder's normalization and the projection of the phase-shift values into unit modulus values are handled in independent normalization/projection layers. This structure enables high flexibility in power allocation since individual constraints can be assigned to the users. Besides, it is an appealing choice for multi-agent implementations since the backbone network can be trained centrally and shared with all the users and the \ac{IRS} controller. The hidden layers are fully connected layers. They are composed of $2D_\text{state}^{\text{DCB}}$ output neurons in the network backbone, $N_\textsubscript{t}N_\textsubscript{s}$ neurons in each normalization head, and $N$ neurons in the projection head. We made several tests with different configurations, and no improvement was observed by increasing this number. We use the \ac{ReLU} function as the activation function in the layers within the network backbone, whereas a linear function was employed in the multiple heads. We include a final concatenation layer to construct the action vector from the multiple heads.

The dimension of the input layer in the critic network equals $D_\text{state}^{\text{DCB}}+D_\text{action}^{\text{DCB}}$ since it takes both state and action vectors as inputs. As shown in \cref{DCB_ANN}, these vectors are first treated independently and later concatenated inside the network. The hidden layers are also fully connected with $2(D_\text{state}^{\text{DCB}}+D_\text{action}^{\text{DCB}})$ neurons. Similarly, we use the \ac{ReLU} activation function in the hidden layers. The output layer dimension is one since this network aims to predict the value of the reward function for a given duple of state and action. Hence, we use the linear activation function at this output layer. Finally, we use the Adam optimizer in both networks since this algorithm has proven to be computationally efficient and robust for supervised and deep reinforcement learning problems \cite{huang_reconfigurable_2020, feng_deep_2020, kingma_adam_2017}.

\section{RL-based Joint Optimization}
In this section, we propose an \ac{RL}-based approach to solving the optimization problem in \eqref{OpA}. In the same way as for \ac{CB}, we start addressing the selection of the state, action, and reward spaces. Next, we will review some important components in the \ac{DRL}-based \ac{DDPG} framework, and finally, we will derive the algorithmic solution. 

\subsection{DRL-DDPG: State, Action and Reward}
As previously explained, the state, action, and reward spaces in \ac{RL} must meet the \ac{MDP} property, i.e., actions will affect the following states as well as the rewards. Besides, states must include information about all the aspects of the past agent-environment interactions that make a difference in the future. Considering this idea, we have selected these elements in the following manner:

\begin{itemize}
\item{The \textbf{state} vector $\mathbf{s}_{t}$ is composed of the current values of all the precoders ($\mathbf{P}_k,\forall k$), the entries in the main diagonal of the \ac{IRS} phase-shift matrix ($\boldsymbol{\theta}$), and entries in the channel response matrices (${\mathbf{{H}}_{\text{IB}}}$ and ${\mathbf{{H}}_{\text{UI}}}_{k},\forall k$). Thus, the state vector is constructed such that

\begin{align}\label{stateDRL}
\mathbf{s}_{t}=[\operatorname{flatten}(\mathbf{P}_1),\ldots,\operatorname{flatten}(\mathbf{P}_K),\operatorname{flatten}(\boldsymbol{\theta}),\\\operatorname{flatten}(\mathbf{H}_\text{IB}), \operatorname{flatten}({\mathbf{{H}}_{\text{UI}}}_{1}),\ldots,\operatorname{flatten}({\mathbf{{H}}_{\text{UI}}}_{K})].\notag
\end{align}

Hence, the dimension of the state space vectors is $D_\text{state}^{\text{DRL}}=KN_\text{t}N_\text{s}+N+N_\text{r}N+KN_\text{t}N$. Recall that the state space is continuous since the entries of $\mathbf{s}_{t}$ can take any complex value.}  

\item{The \textbf{action} vector $\mathbf{a}_{t}$ comprises the matrices $\Delta\boldsymbol{\theta}$ and $\Delta\mathbf{P}_k, \forall k$, whose entries stand for the variations in the values of $\boldsymbol{\theta}$ and $\mathbf{P}_k, \forall k$, respectively. Hence, the dimension of the action space is $D_\text{action}^{\text{DRL}} = KN_\text{t}N_\text{s}+N$ and the action is constructed such that 
\begin{align}\label{actionDRL}
\mathbf{a}_{t}=[\operatorname{flatten}(\Delta\mathbf{P}_1),\ldots,\operatorname{flatten}(\Delta\mathbf{P}_K),\\\operatorname{flatten}(\Delta\boldsymbol{\theta})].\notag
\end{align}
The action space is also continuous. However, its values are constrained to have imaginary and real parts within the real-valued interval $(-1,1)$ to improve the system's stability.}

\item{The \textbf{reward} $r_{t}$ is determined as a function of the sum-rate, which is the metric we aim to maximize. In this DRL-DDPG formulation, rather than using the sum-rate value itself, we calculate $r_{t}$ as the difference between the values after and before taking the action $\mathbf{a}_{t}$, i.e.,  

\begin{align}\label{rewardDRL}
{r}_{t}=\sum_{k=1}^{K}~{R_{k}}(\mathbf{s}_{t}, \mathbf{a}_{t}) - \sum_{k=1}^{K}~{R_{k}}(\mathbf{s}_{t-\text{1}}, \mathbf{a}_{t-\text{1}}).
\end{align}}
\end{itemize}

In the way we have defined them, actions allow us to identify a correspondence between the current state and the action taken, and the corresponding next state\textemdash as expected in DRL-based formulations. Furthermore, the immediate reward depends on the state and the action, not only on the action. 

The ANNs used for function approximation are sensitive to the scale of the features. If we use the sum-rate values as rewards, the difference between the scales of actions and rewards can be unfavorable to the learning and the stability of the consecutive time steps. By considering the reward to be the difference between the sum-rate values, the scales of state, action, and reward values remain similar.

\subsection{DRL-DDPG: Framework Elements}
\begin{figure}[!t]
	\centering
	\includegraphics[width=0.9\columnwidth]{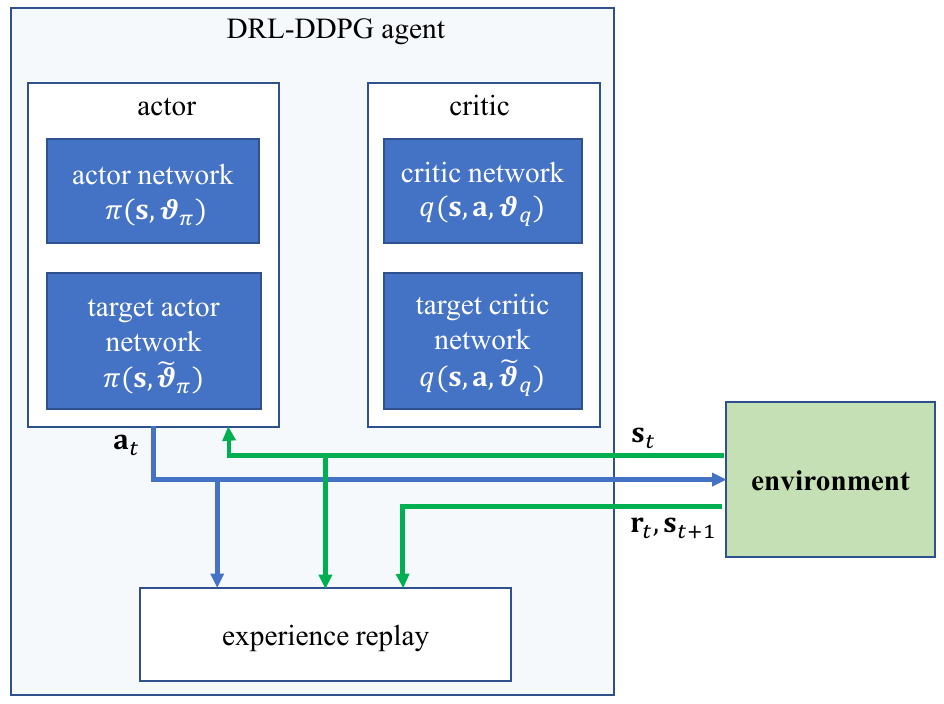}
	\caption{Elements in the DRL-DDPG agent.}
	\label{DRLframework}
\end{figure}

\Cref{DRLframework} shows a representation of the DRL-DDPG agent. As observed, several of these elements were already used in the previously described \ac{DCB-DDPG} framework. The only additional elements required for this framework are the \textbf{target networks}. In this case, the initial parameters of the actor and critic networks are copied into the target actor network $\pi(\mathbf{s},\tilde{\boldsymbol{\vartheta}}_\pi)$ and target critic network $q(\mathbf{s}, \mathbf{a},\tilde{\boldsymbol{\vartheta}}_\text{q})$, respectively. These target networks are used for calculating the target values, and their network parameters ($\tilde{\boldsymbol{\vartheta}}_\pi$ and $\tilde{\boldsymbol{\vartheta}}_\text{q}$) are updated in a soft-manner to track the original networks. Using these target networks makes it possible to set stable targets to improve the convergence of the critic network training. These target networks are not considered in the \ac{DCB} approach since they are related to the estimations of future actions and rewards, which are relevant only to the \ac{DRL} approach. Notice that in this case, we refer to the critic network as $q(\mathbf{s}, \mathbf{a},{\boldsymbol{\vartheta}}_\text{q})$ since it approximates the long-term reward function $q_\pi(\cdot)$.

\subsection{DRL-DDPG Algorithm}
In this section, we present the algorithm proposed to train the \ac{DRL}-\ac{DDPG} agent for computing the \ac{IRS} phase-shift matrix ($\mathbf{\Theta}$) and the precoder ($\mathbf{P}$) that maximize the achievable sum-rate in \eqref{OpA}. \Cref{DRL_alg} summarizes the steps to follow. Along the analysis, we focus on the main differences regarding the \ac{DCB}-\ac{DDPG} algorithm. 

The actor and critic networks are created through the randomly initialized network parameters ($\boldsymbol{\vartheta}_\pi$ and $\boldsymbol{\vartheta}_\text{q}$, respectively). In this DRL-DDPG approach, the target actor and target critic networks are also created as their exact copies. Hence, before the learning process starts, $\boldsymbol{\vartheta}_\pi = \tilde{\boldsymbol{\vartheta}}_\pi$ and $\boldsymbol{\vartheta}_\text{q} = \tilde{\boldsymbol{\vartheta}}_\text{q}$. 

\begin{algorithm}[ht!]
\caption{DRL-DDPG algorithm}\label{DRL_alg}
\begin{algorithmic}[1]
\STATE \hspace{0.1cm}$ \textbf{Initialize: }$
\STATE \hspace{0.1cm}$ \text{set }\pi(\mathbf{s},\boldsymbol{\vartheta}_\pi) \text{ given } \boldsymbol{\vartheta}_\pi \text{, and }q(\mathbf{s}, \mathbf{a},\boldsymbol{\vartheta}_\text{q})\text{ given }\boldsymbol{\vartheta}_\text{q}$
\STATE \hspace{0.1cm}$\pi(\mathbf{s},\tilde{\boldsymbol{\vartheta}}_\pi) \gets \pi(\mathbf{s},\boldsymbol{\vartheta}_\pi) \text{, and } q(\mathbf{s}, \mathbf{a},\tilde{\boldsymbol{\vartheta}}_\text{q})\gets q(\mathbf{s}, \mathbf{a},\boldsymbol{\vartheta}_\text{q})$ 
\STATE \hspace{0.1cm}$ \text{create } \mathcal{R}$
\STATE \hspace{0.1cm}$ \textbf{for }e = 0,\ldots, E-1 \textbf{ do:} $
\STATE \hspace{0.5cm}$ \text{set }\mathbf{s}\textsubscript{0} \text{ given random }\mathbf{P} \in \mathcal{P}\text{ and } \mathbf{\Theta}\in \mathcal{D}  $
\STATE \hspace{0.5cm}$ \textbf{for }t = 0,\ldots, T-1\textbf{ do:} $
\STATE \hspace{1cm}$ \text{agent takes }\mathbf{a}_{t} = \pi(\mathbf{s}_{t},\boldsymbol{\vartheta}_\pi)+ \mathbf{n}_\text{e}$
\STATE \hspace{1cm}$ \text{environment returns }r_{t} \text{ and }\mathbf{s}_{t+1}$
\STATE \hspace{1cm}$ \mathcal{R} \text{ stores } \mathcal{E}_t = (\mathbf{s}_{t}, \mathbf{a}_{t}, r_{t},
\mathbf{s}_{t+1})$
\STATE \hspace{1cm}$ \textbf{if }|\mathcal{R}| > |\mathcal{B}|:$
\STATE \hspace{1.5cm}$ \text{agent samples }|\mathcal{B}|\text{ random experiences: }$
\STATE \hspace{1.5cm}$ \mathcal{E}_i=(\mathbf{s}_{i}, \mathbf{a}_{i}, r_{i}, \mathbf{s}_{i+1}), ~i=0,\ldots, |\mathcal{B}|-1 $
\STATE \hspace{1.5cm}$ \text{calculate }\tilde{\mathbf{a}}_{i+1}=\pi(\mathbf{s}_{i+1},\tilde{\boldsymbol{\vartheta}}_\pi)$
\STATE \hspace{1.5cm}$ \text{calculate }\tilde{y}_{i}=r_i+\gamma q(\mathbf{s}_{i+1}, \tilde{\mathbf{a}}_{i+1},\tilde{\boldsymbol{\vartheta}}_\text{q})$
\STATE \hspace{1.5cm}$ \text{calculate the critic loss } L_{\text{c}} \text{ by using \eqref{LcDRL}} $
\STATE \hspace{1.5cm}$ \text{update } \boldsymbol{\vartheta}_\text{q} \text{ by the back-propagation of }L_{\text{c}}$
\STATE \hspace{1.5cm}$ \text{calculate the actor loss } L_{\text{a}} \text{ by using \eqref{LaDRL}} $
\STATE \hspace{1.5cm}$ \text{update } \boldsymbol{\vartheta}_\pi \text{ by the back-propagation of }L_{\text{a}}$
\STATE \hspace{1.5cm}$ \tilde{\boldsymbol{\vartheta}}_\pi \gets \tau \boldsymbol{\vartheta}_\pi + (1-\tau)\tilde{\boldsymbol{\vartheta}}_\pi $
\STATE \hspace{1.5cm}$ \tilde{\boldsymbol{\vartheta}}_\text{q} \gets \tau \boldsymbol{\vartheta}_\text{q} + (1-\tau)\tilde{\boldsymbol{\vartheta}}_\text{q}$
\STATE \hspace{1cm}$ \textbf{end if }$
\STATE \hspace{0.5cm}$ \text{Obtain } \mathbf{P} \text{ and } \mathbf{\Theta} \text{ by evaluating policy }\pi(\mathbf{s},\boldsymbol{\vartheta}_\pi)$
\end{algorithmic}
\textbf{Output: }\text{trained actor network }$\pi(\mathbf{s},\boldsymbol{\vartheta}_\pi)$
\end{algorithm}

The interactions between the DRL-DDPG agent and the environment are split into $E$ episodes with a finite number of time steps $T$ since no terminal states can be defined in our problem. Dividing the training time steps into episodes enables the learning process to start from different initial states, thus allowing exploration over the entire state space. We consider setting $T = 20$ and varying the number of episodes according to the intended number of interactions. 

The initial state $\mathbf{s}_{0}$ is created as in \eqref{stateDRL} from the randomly initialized $\mathbf{P}$ and $\mathbf{\Theta}$ matrices, and the randomly generated channel response matrices (${\mathbf{{H}}_{\text{UI}}}_{k},\;\forall k$ and ${\mathbf{{H}}_{\text{IB}}}$). Starting from that state, the agent takes actions according to the output of the actor network and an exploration noise $\mathbf{n}_\text{e}\sim \mathcal{N}_{\mathbb{C}}(0, \sigma_{\text{n}_\text{e}}^2\B{I}_{D_\text{action}^{\text{DRL}}})$. As in the DCB-DDPG approach, the exploration noise guarantees a better exploration of the entire action space. Besides, in this DRL-DDPG approach, exploration noise also improves the exploration over the state space because actions affect the occurrence of the following states.

At each learning time step, the experience tuples in $\mathcal{B}$ are employed to train the actor and critic networks as described in lines 12 to 19. As presented in \cite{lillicrap_continuous_2019}, the target variables $\tilde{\mathbf{a}}_{i+1}$ and $\tilde{y}_{i}$ are computed using the target actor network and the target critic network, respectively (lines 14 and 15). The factor $\gamma$ in the computation of $\tilde{y}_{i}$ stands for a discount rate and determines the relevance of future rewards. The critic loss $L_{\text{c}}$ is next calculated as
\begin{align}\label{LcDRL}
L_{\text{c}} = \frac{1}{|\mathcal{B}|}\sum_{i} (\tilde{y}_{i}-q(\mathbf{s}_{i}, \mathbf{a}_{i},\boldsymbol{\vartheta}_\text{q}))^2,
\end{align} 
and the obtained values are back-propagated by using the critic network optimizer. This way, the critic network is trained to minimize $L_{\text{c}}$ and approximate the actual behavior of the long-term reward function $q_\pi(\cdot)$.

On the other hand, the actor network is trained to predict the action that maximizes the output of the critic network. Hence, the actor loss $L_{\text{a}}$ is calculated as
\begin{align}\label{LaDRL}
L_{\text{a}} = -\frac{1}{|\mathcal{B}|}\sum_{i} q(\mathbf{s}_{i}, \pi(\mathbf{s}_{i},\boldsymbol{\vartheta}_\pi),\boldsymbol{\vartheta}_\text{q}),
\end{align} 
and the obtained values are then utilized to update the parameters of the actor network.

The actor and critic target networks are updated as described in lines 20 and 21. Parameter $\tau$ defines the updating rate for the target networks. At the end of every learning time step, the obtained policy is evaluated by running an evaluation episode. We do not consider the exploration noise or update the networks during this stage. By following the actor network policy, the randomly initialized $\mathbf{P}$ and $\mathbf{\Theta}$ matrices are transformed to improve the sum-rate of the system.

\subsection{DRL-DDPG: \acp{ANN} Structure} 

\begin{figure}[!t]
	\centering
	\includegraphics[width=0.8\columnwidth]{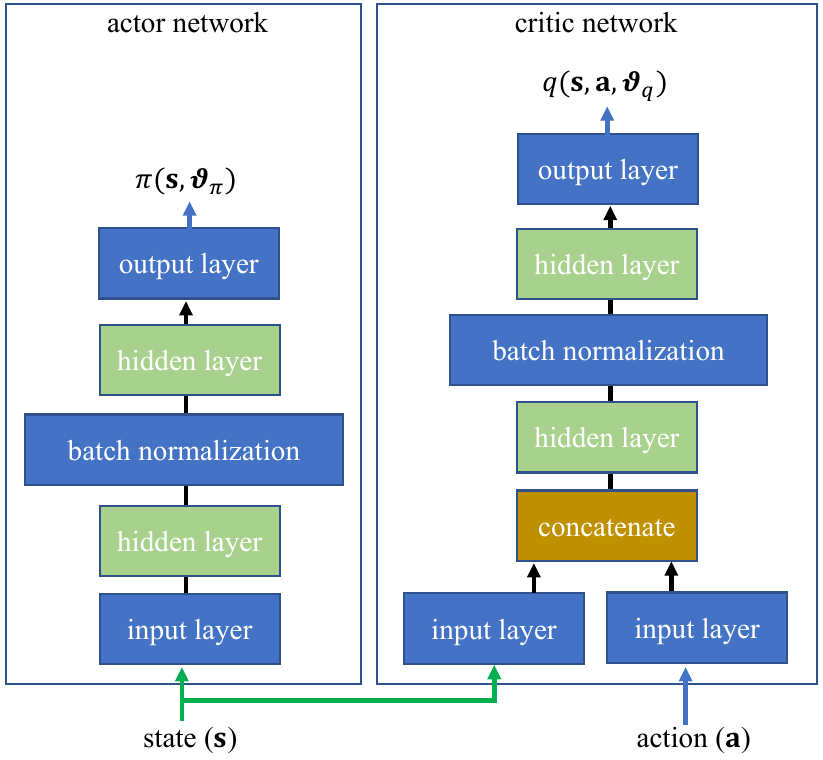}
	\caption{DRL-DDPG actor and critic network structure.}
	\label{DRL_ANN}
\end{figure}

The structures of the \acp{ANN} we propose to use as actor and critic networks in DRL-DDPG are presented in \cref{DRL_ANN}. The input and output layer dimensions of the actor network equal $D_\text{state}^{\text{DRL}}$ and $D_\text{action}^{\text{DRL}}$, respectively. Two fully connected hidden layers with $2D_\text{state}$ neurons were considered for this network. As proposed in \cite{lillicrap_continuous_2019, ioffe_batch_2015, huang_reconfigurable_2020}, we include batch normalization layers in both actor and critic networks to improve the learning performance. By inserting batch normalization layers, higher learning rates can be considered, and the network initialization values have a lower impact on the network performance. We used the \ac{ReLU} function at both hidden layers, while the \ac{tanh} activation function was used at the output layer. 

The critic network input and output dimensions equal $D_\text{state}^{\text{DRL}}+D_\text{action}^{\text{DRL}}$ and 1, respectively. The first layer, in this case, is a concatenation layer that transforms the state and action vectors into one. The inner structure of this network is similar to the actor but considers $2(D_\text{state}^{\text{DRL}}+D_\text{action}^{\text{DRL}})$ neurons at the hidden layers. The single-neuron output layer uses the linear activation function to predict the long-term reward value for the action and state inputs.

\section{Convergence and Computational Complexity}
In this section, we analyze the convergence and computational complexity of the proposed approaches. We make a brief comparison to identify how learning and computing costs are affected by the framework structures and some configuration parameters.    

\subsection{Computational Complexity}
In the following, we present a computational complexity analysis based on the number of required multiplications. As stated in \cite{freire_computational_2022}, this is a high-level metric for computational complexity evaluation because only multiplications are considered while other less time-consuming operations are disregarded. We consider two stages that occur at different moments and have different resource limitations. First, we address the computational complexity of the training stages (described in \Cref{DCB_alg} and \Cref{DRL_alg}), which are expected to occur mostly offline. Second, we address the computational complexity of the operational stages where the already trained actor networks are used to predict the best IRS phase-shift and precoder matrices for the observed channel realizations.

Let us start by analyzing the computational complexity of the training stage in the DCB-DDPG approach (\Cref{DCB_alg}). Since training can be performed for as many time steps as desired, we consider only the operations made within a time step in both approaches. The highest computational complexity in \Cref{DCB_alg} is in the operations with the critic network. When considering mini-batches of size $|\mathcal{B}|$, the computational complexity for \acp{ANN} with fully connected layers is bounded by $\mathcal{O}\left(|\mathcal{B}|\psi\zeta^2\right)$, where $\psi$ is the number of hidden layers, and $\zeta$ is the number of neurons in the widest layer \cite{freire_computational_2022,zhao_dynamic_2021}. We disregard the number of layers $\psi$ because it is not related to the communication parameters, and its value is generally small compared to $|\mathcal{B}|$. The worst condition for $\zeta$ is considering the hidden layers of the critic network, which contain $2(D_\text{state}^{\text{DCB}}+D_\text{action}^{\text{DCB}})$ neurons. Hence, the computational complexity of this step, and the algorithm in general, is in the order of $\mathcal{O}\left(|\mathcal{B}|\left(K^2N_\text{t}^2N^2+N_\text{r}^2N^2+K^2N_\text{t}^2N_\text{s}^2+N^2\right)\right)$. 

The computational complexity analysis for the training stage in DRL-DDPG is similar. In \Cref{DRL_alg}, the highest complexity also corresponds to the calculations over the critic network. Hence, following the same reasoning, the computational complexity of this algorithm is bounded by $\mathcal{O}\left(|\mathcal{B}|\left(K^2N_\text{t}^2N^2+N_\text{r}^2N^2+4K^2N_\text{t}^2N_\text{s}^2+4N^2\right)\right)$. As observed, the computational cost of training in DRL-DDPG is higher due to the larger number of elements being considered in the states. Notice that, beyond the increment of the computational complexity, larger state vectors also imply larger storage capabilities needed for the \ac{ANN} models and the replay buffer.

In the operational stage, the computational complexity is expected to be remarkably lower, which is desirable to fit the latency requirements of practical applications. In this stage, no exploration is performed, but experiences are stored for offline training. In the DCB-DDPG approach, the already trained actor network is used to predict the best IRS phase-shift and precoder matrices for a given channel realization in a single-step forward pass. As explained in Section IV, the widest layers in the DCB-DDPG actor network are in the backbone, with $2D_\text{state}^{\text{DCB}}$ neurons. Hence, the computational complexity of finding the best matrices in DCB-DDPG is in the order of $\mathcal{O}\left(K^2N_\text{t}^2N^2+N_\text{r}^2N^2\right)$. 

On the other hand, in the DRL-DDPG approach, the IRS phase-shift and precoder matrices are predicted through the sequence of $T$ consecutive time steps that form an episode. In this case, the widest layers in the actor network for DRL-DDPG are composed of $D_\text{state}^{\text{DRL}}$ neurons. Hence, the computational complexity of the operational stage with DRL-DDPG is in the order of $\mathcal{O}\left(T\left(K^2N_\text{t}^2N^2+N_\text{r}^2N^2+K^2N_\text{t}^2N_\text{s}^2+N^2\right)\right)$.

As shown before, the DCB-DDPG approach outperforms the DRL-DDPG scheme in terms of computational complexity. By presenting simpler implementations of the \ac{ANN}-based function approximations, and a simpler algorithm, DCB-DDPG stands as a more suitable alternative to handle the joint optimization problem, especially in applications with stringent latency requirements or computing resource constraints.

\subsection{Convergence} 

As explained in \cite{sutton_reinforcement_2018}, convergence cannot be guaranteed in algorithms like those proposed in DCB-DDPG and DRL-DDPG. These algorithms combine three crucial features, which are known as \textit{the deadly triad}: function approximation, bootstrapping (i.e., using previous estimates for training), and off-policy training. They are called this way because instability and divergence arise when these features are jointly employed. However, none of these features can be given up because of their relevance. For this reason, the convergence analysis must be empirically addressed by considering the specific conditions of our frameworks and system model. 

\begin{figure}[!t]
	\centering
	\includegraphics[width=1\columnwidth]{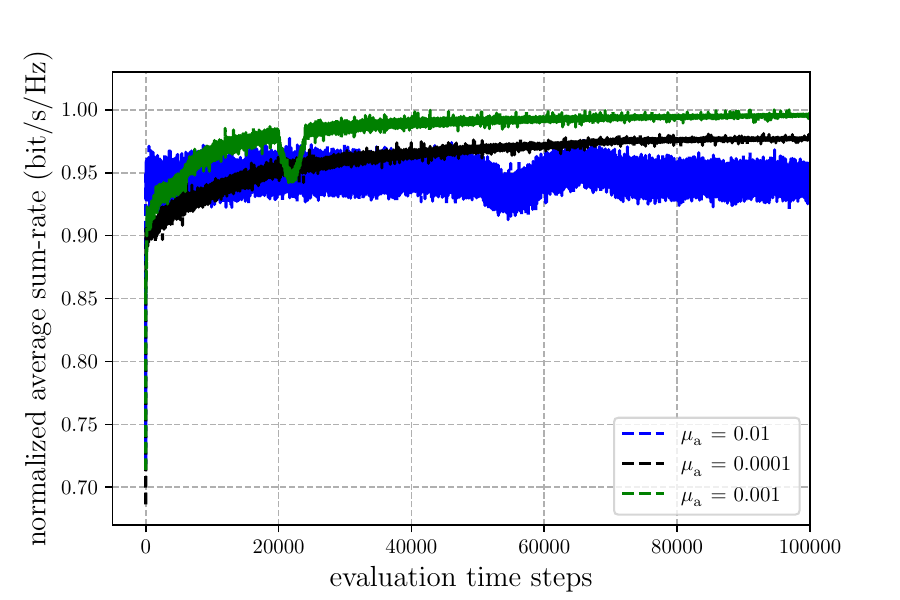}
	\caption{Normalized average sum-rate (bit/s/Hz) vs evaluation time steps with the DCB-DDPG framework.}
	\label{aveSR_LR}
\end{figure}

\begin{figure}[!t]
	\centering
	\includegraphics[width=1\columnwidth]{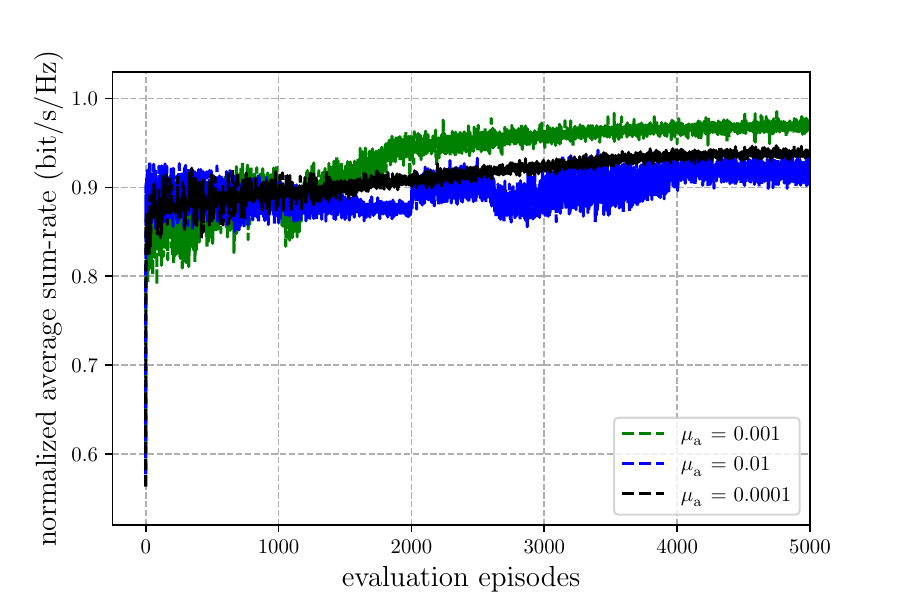}
	\caption{Normalized average sum-rate (bit/s/Hz) vs evaluation episodes with the DRL-DDPG framework.}
	\label{aveSR_LR_RL}
\end{figure}

As in \cite{yang_deep_2021, stylianopoulos_deep_2022, huang_reconfigurable_2020, xu_experience-driven_2021, zhao_dynamic_2021}, we first propose a convergence analysis based on the reward function. \Cref{aveSR_LR} and \Cref{aveSR_LR_RL} show the normalized average sum-rate values computed at the evaluation time steps in DCB-DDPG and the evaluation episodes in DRL-DDPG, respectively. At each evaluation stage, the average sum-rate values are computed over 100 channel realizations different from those visited during training. Normalization is performed by considering the overall highest average sum-rate value as the normalization factor to allow a fair comparison. These figures show that convergence in terms of sum-rate can be empirically illustrated for both approaches in the considered setup. The obtained values steadily improve and converge to their best solutions when the learning rates of the actor network optimizers ($\mu_{\textsubscript{a}}$) are equal to 0.001. Thus, the proper behavior of the actor networks is demonstrated, since they continuously improve on learning the best precoder and the \ac{IRS} phase-shift matrix for each channel realization. 

\begin{figure}[!t]
	\centering
	\includegraphics[width=1\columnwidth]{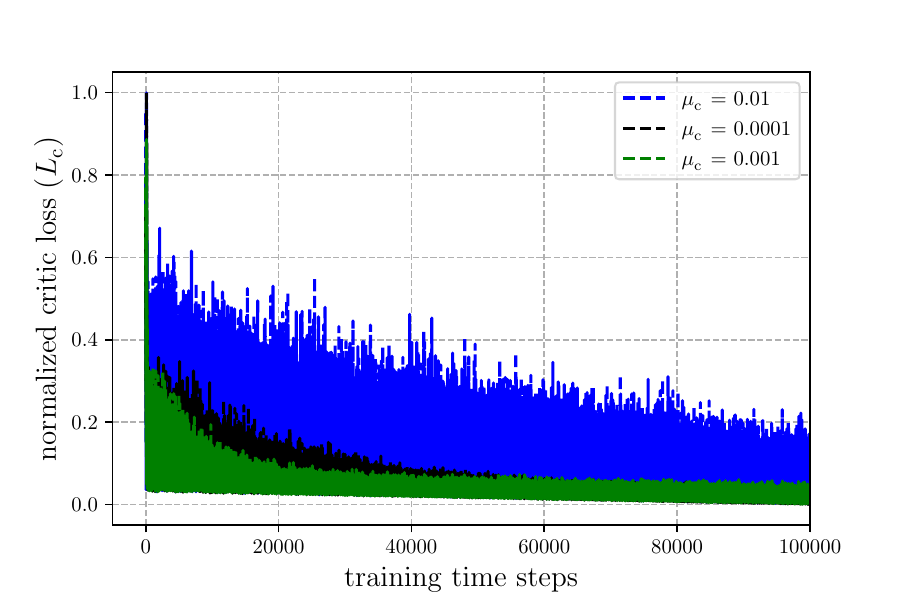}
	\caption{Normalized critic loss vs training time steps with the DCB-DDPG framework.}
	\label{criticLoss_LR}
\end{figure}

\begin{figure}[!t]
	\centering
	\includegraphics[width=1\columnwidth]{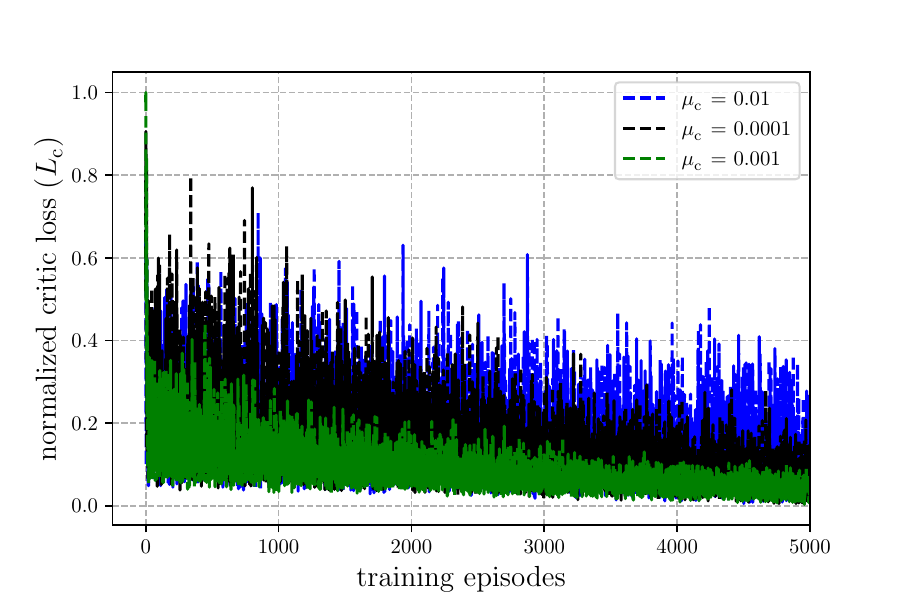}
	\caption{Normalized critic loss vs training episodes with the DRL-DDPG framework.}
	\label{criticLoss_LR_DRL}
\end{figure}

Additionally, we have performed a second convergence analysis based on the critic loss values ($L_{\text{c}}$). The better the critic networks approximate the instantaneous reward function $r_t(\cdot)$ in DCB-DDPG and the long-term reward function $q_\pi(\cdot)$ in DRL-DDPG, the smaller the values of $L_{\text{c}}$ are. \Cref{criticLoss_LR} and \Cref{criticLoss_LR_DRL} show the normalized critic loss values in the DCB-DDPG and the DRL-DDPG approaches, respectively. As in the previous analysis, the convergence of both algorithms can be empirically shown for at least one of the learning rates of the critic network optimizer  ($\mu_{\textsubscript{c}}=0.001$). Notice that, in both convergence analyses, higher learning rate values ($\mu_{\textsubscript{a}}=\mu_{\textsubscript{c}}=0.01$) lead to lower stability. On the other hand, lower learning rates ($\mu_{\textsubscript{a}}=\mu_{\textsubscript{c}}=0.0001$) improve stability but sacrifice convergence speed.

As observed in the previous figures, the proposed DCB-DDPG framework is more robust regarding sum-rate maximization and critic loss minimization. In both cases, DCB-DDPG converges to better solutions while keeping lower variance values. In this sense, convergence speed and stability are mainly related to the complexity of the learning and evaluation stages. In DCB-DDPG, single-time step predictions are used to find the best precoder and IRS phase-shift matrices. Hence, either good or bad, predictions in DCB-DDPG only affect the instantaneous reward and not the future states or rewards. On the other hand, matrices in DRL-DDPG are predicted through the sequence of consecutive time steps that form an episode. Hence, prediction errors might propagate and increase from one time step to the next, thus reducing the system stability.

\section{Simulation Results}

In this section, we present results that validate the use of the \ac{DCB-DDPG} and DRL-DDPG frameworks to optimize the IRS phase-shift and precoder matrices in the uplink of an \ac{IRS}-assisted \ac{MS} MU-MIMO system, as the one described in Section III. We consider the setups presented in Sections IV and V, and follow the steps described in \Cref{DCB_alg} and \Cref{DRL_alg}, respectively.  

\Cref{tableSimul} shows the configuration parameters considered for the computer experiments. For fairness, we have selected the number of episodes in DRL-DDPG and the time steps in both proposals so that the training times remain similar for both approaches. In both cases, the mini-batches with 16 entries constitute an adequate trade-off between learning performance and complexity. The learning rate values in both proposals and the update and discount factors in DRL-DDPG were obtained experimentally through a grid search approach to provide the best performance in terms of the system's achievable sum-rate. The exploration noise variance was set to 0.05 in both approaches to guarantee a fair trade-off between exploration and exploitation over the action space. This value is also in concordance with the one proposed in \cite{lillicrap_continuous_2019}. As explained, exploration over state space is also guaranteed by the randomness of channel realizations and the episodes in the DRL-DDPG approach.

\begin{table}[!t]
\caption{Configuration Parameters\label{tableSimul}}
\centering
\begin{tabular}{|c|c|c|}
\hline
\multicolumn{3}{|c|}{DCB-DDPG parameters}  \\
\hline
Parameter &Description & Value\\
\hline
$T$ & number of training time steps & 100000\\
\hline
$|\mathcal{B}|$ & mini-batch size  & 16\\
\hline
$|\mathcal{R}|_{\text{max}}$ & max. experience replay buffer size  & 100000\\
\hline
$\mu_{\text{a}}$ & actor network optimizer learning rate  & 0.001\\
\hline
$\mu_{\text{c}}$ & critic network optimizer learning rate  & 0.001\\
\hline
$\sigma_{\text{n}_\text{e}}^2$ & exploration noise variance  & 0.05\\
\hline
\hline
\multicolumn{3}{|c|}{DRL-DDPG parameters}  \\
\hline
Parameter &Description & Value\\
\hline
$E$ & number of episodes & 5000\\
\hline
$T$ & number of time steps per episode  & 20\\
\hline
$|\mathcal{B}|$ & mini-batch size  & 16\\
\hline
$|\mathcal{R}|_{\text{max}}$ & max. experience replay buffer size  & 100000\\
\hline
$\mu_{\text{a}}$ & actor network optimizer learning rate  & 0.001\\
\hline
$\mu_{\text{c}}$ & critic network optimizer learning rate  & 0.001\\
\hline
$\tau$ & target network updating rate  & 0.005\\
\hline
$\gamma$ & discount factor  & 0.99\\
\hline
$\sigma_{\text{n}_\text{e}}^2$ & exploration noise variance  & 0.05\\
\hline
\end{tabular}
\end{table}

We have selected two model-driven approaches as benchmarks. The first scheme, termed Alternating-ProG, uses an alternating minimization projected gradient algorithm to optimize the IRS and precoder matrices \cite[Algorithm 5]{perovic_maximum_2021}. This solution has been shown to perform better than other strategies proposed in the literature for the joint design of the IRS and precoding matrices.
We have also included a baseline strategy termed RandomIRS-MRT to provide a lower-bound reference. In the RandomIRS-MRT scheme, the IRS phase-shift matrix is randomly selected from the set of feasible matrices $\mathcal{D}$, and the precoders are computed according to the \ac{MRT} criterion.

\begin{figure}[!t]
	\centering
	\includegraphics[width=1\columnwidth]{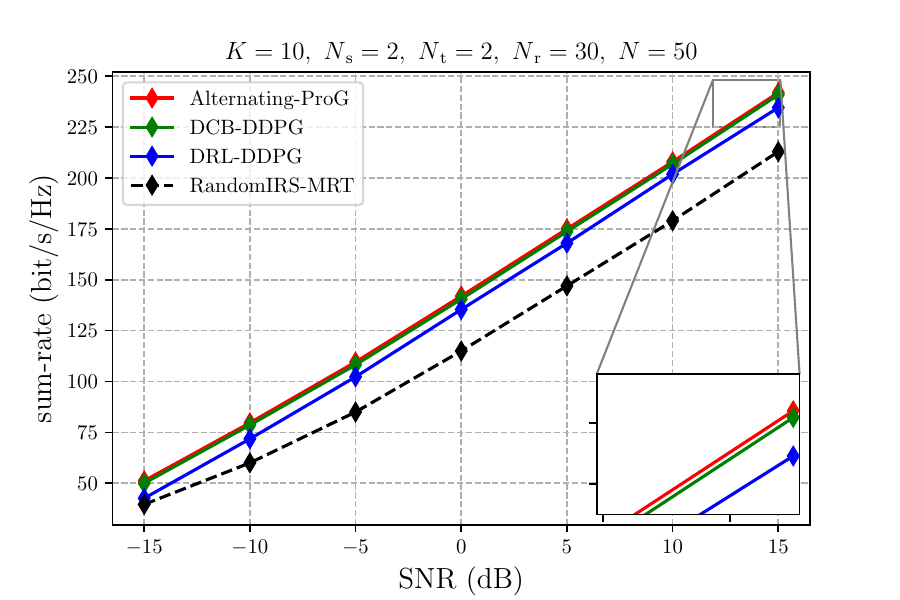}
	\caption{Sum-rate (bit/s/Hz) vs SNR (dB) for $K=10, N\textsubscript{s}=2, N\textsubscript{t}=2,N\textsubscript{r}=30,N=50$.}
	\label{MS_SR_SNR}
\end{figure}

\Cref{MS_SR_SNR} shows the achievable sum-rates obtained with the proposed \ac{DCB-DDPG} and DRL-DDPG approaches and the two benchmarks for a range of \ac{SNR} values between -15 dB and 15 dB. The results in \Cref{MS_SR_SNR} were obtained by considering a setup with $K=10$ users employing $N_\text{t}=2$ antennas to send $N_\text{s}=2$ streams each, an \ac{IRS} with $N=50$ scattering elements, and a \ac{BS} with $N_\text{r}=30$ receiving antennas. 

In these simulation conditions, the Alternating-ProG benchmark outperforms our proposals, although the difference concerning the \ac{DCB-DDPG} scheme is minimal. This is an expected result that an important remark can explain. As stated in \cite{bjornson_massive_2017}, the performance of linear algorithms is near optimal when $N_\text{r}>>KN_\text{s}$, as occurs in this configuration. Although the necessary alternating procedure imposes certain limitations on its performance, we can say that the communication configuration is favorable for the benchmark approach. The performance of DRL-DDPG in this scenario is also good, although it is outperformed by the DCB-DDPG algorithm. Since we constrained the training times of both approaches to be equal, the higher complexity of DRL-DDPG and better suitability of the \ac{DCB} formulation for this problem led to the observed gap in their performances. On the other hand, both proposals significantly outperform the RandomIRS-MRT baseline strategy. 

\begin{figure}[!t]
	\centering
	\includegraphics[width=1\columnwidth]{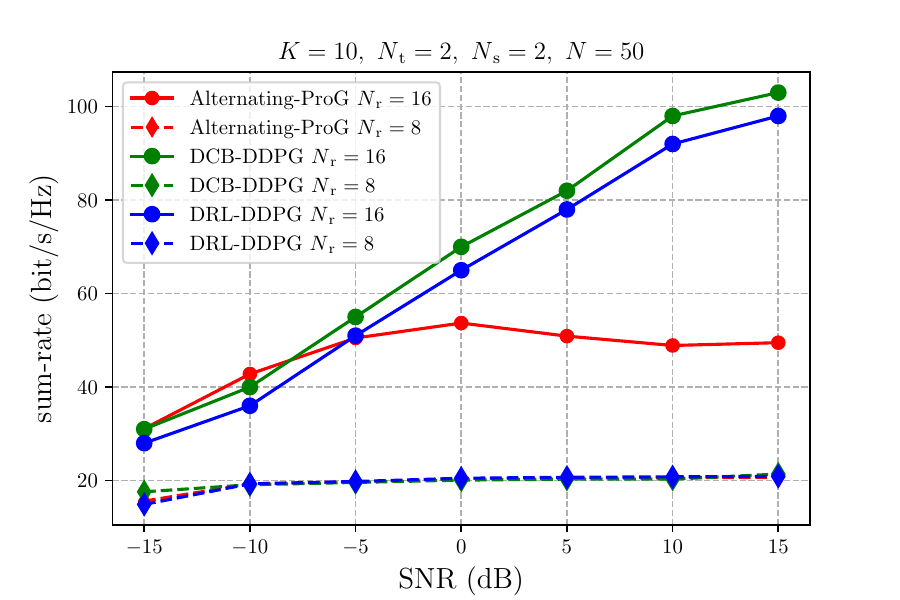}
	\caption{Sum-rate (bit/s/Hz) vs SNR (dB) for $N\textsubscript{r}\in{\lbrace8,16\rbrace},K=10, N\textsubscript{t}=2, N\textsubscript{s}=2, N=50$.}
	\label{MS_SR_Comp}
\end{figure}

We next considered the same setup as in the previous experiment and conducted simulations for two different numbers of \ac{BS} receiving antennas, namely $N_\text{r} \in \{8,16\}$. The results are shown in \Cref{MS_SR_Comp}. Note that for $N_\text{r}=16$, the total number of transmitted data streams $KN_\text{s}=20$ is slightly higher than the number of receiving antennas $N_\text{r}$. In the low SNR regime (SNR $<$ -5 dB), the behavior of the sum-rate values is similar to that observed in \Cref{MS_SR_SNR}. The \ac{DCB-DDPG} and Alternating-ProG schemes behave similarly and slightly outperform DRL-DDPG. This result is reasonable because, in the low \ac{SNR} regime, the system performance is more sensitive to channel noise than to multi-user interference. However, when SNR increases, Alternating-ProG is unable to handle the multi-user interference, and the achievable sum-rate decreases or remains constant. On the other hand, both DCB-DDPG and DRL-DDPG manage the interference more appropriately, and the sum-rate increases almost linearly with the SNR values. For example, the achievable sum-rate obtained with DCB-DDPG when $\text{SNR} = 15 \text{ dB}$ is more than 40 bits/s/Hz higher than that provided by Alternating-ProG. In addition, DCB-DDPG outperforms DRL-DDPG by around 5 bits/s/Hz in all the SNR regimes.

\Cref{MS_SR_Comp} also shows the results obtained when $N_\text{r}=8$. This is a highly demanding setup since the total of transmitted data streams $KN_\text{s}=20$ is higher than twice the number of receiving antennas. Under this condition, none of the algorithms is able to tackle the multi-user interference, and a scheduling stage should be considered.

\begin{figure}[!t]
	\centering
	\includegraphics[width=1\columnwidth]{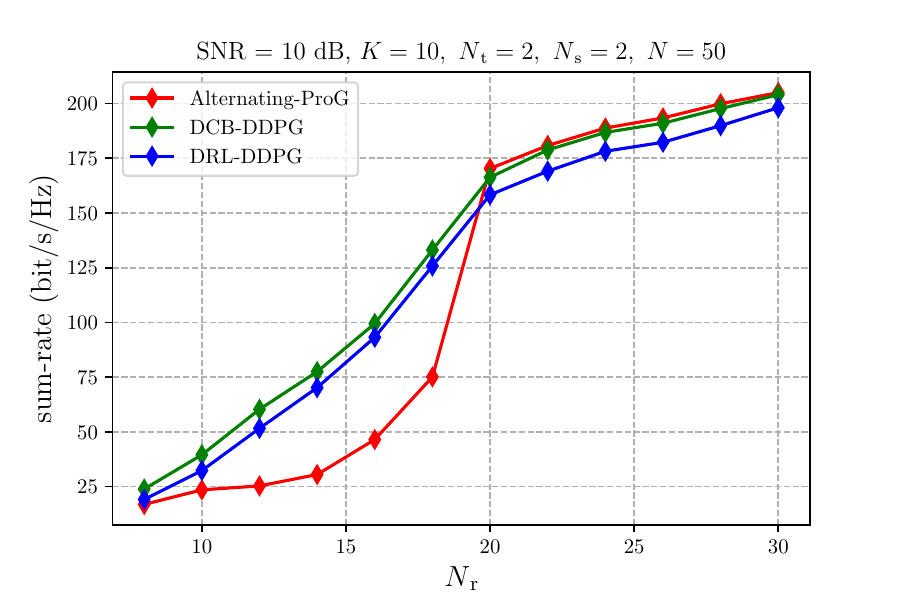}
	\caption{Sum-rate (bit/s/Hz) vs $N_\text{r}$ for SNR = 10 dB,$K=10, N\textsubscript{t}=2, N\textsubscript{s}=2, N=50$.}
	\label{MS_SR_Nr}
\end{figure}

We conducted another set of experiments to analyze the performance impact of the relationship between the number of receiving antennas $N_\text{r}$ and the number of transmitted streams $KN_\text{s}$. We again assumed a setup with $K=10$, $N_{\text{s}}=2$, $N_{\text{t}}=2$ and $N=50$. We fixed $\text{SNR}=10 \text{ dB}$ and considered $N_{\text{r}}$ values that range from 8 to 30. \Cref{MS_SR_Nr} shows the obtained results. In the following, we analyzed them by considering three working regimes: $N_\text{r}/KN_\text{s}<<1$, $N_\text{r}/KN_\text{s}<1$, and $N_\text{r}/KN_\text{s}\geq1$. 

The first regime with $N_\text{r}/KN_\text{s}<<1$ represents a critical scenario where, as discussed previously, none of the analyzed algorithms perform well, and their sum-rates degrade remarkably. In the third regime with $N_\text{r}/KN_\text{s}\geq1$, all the algorithms perform reasonably well, although DCB-DDPG and Alternating-ProG outperform DRL-DDPG. It is also worth noting that the DCB-DDPG scheme nearly achieves the same performance as the model-driven benchmark in this regime. 
This behavior matches that observed in \Cref{MS_SR_SNR}. The second regime with $N_\text{r}/KN_\text{s}<1$ is an intermediate scenario where DCB-DDPG and DRL-DDPG clearly outperform Alternating-ProG. When moving into this regime, Alternating-ProG cannot properly handle the interference among the users. The sum-rate values slowly increase when considering more receiving antennas, and no major performance improvement is observed until the third regime condition is met. On the other hand, DCB-DDPG and DRL-DDPG manage the interference more efficiently, and the sum-rate values increase almost linearly when considering more receiving antennas. When $N_{\text{r}}$ decreases from 20 to 18, the Alternating-ProG sum-rate value degrades near 75 bit/s/Hz, while this value only degrades around 25 bit/s/Hz for DCB-DDPG and DRL-DDPG. The results show that the sum-rate values achieved with the proposed schemes are at least 25 bit/s/Hz higher when considering $N_{\text{r}}$ values between 12 and 18.

As explained in Section II, formulating the joint optimization problem as \ac{RL} or \ac{CB} with discrete actions is not feasible. The number of possible actions becomes intractable even for low-complexity setups. The number of possible actions\textemdash and the size of the output layer of the \ac{ANN} used for the policy function approximation\textemdash would be as large as $\approx 10^{15}$, only by addressing the optimization of an \ac{IRS} with 50 scattering elements and two possible phase values. This number is several times larger if we also consider discrete-valued precoding vectors. However, to meet the requirements of practical implementations, we have analyzed the use of an additional discretization stage to be applied to the already computed optimal continuous-valued \ac{IRS} matrices, instead of considering discrete formulations of the problem.

\begin{figure}[!t]
	\centering
	\includegraphics[width=1\columnwidth]{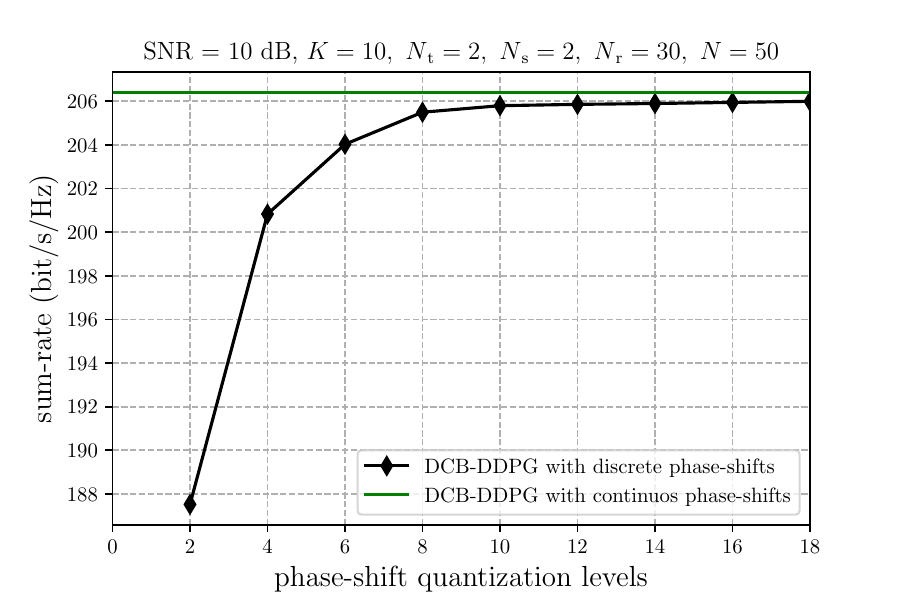}
	\caption{Sum-rate (bit/s/Hz) vs \ac{IRS} phase-shift matrix quantization levels.}
	\label{quantization}
\end{figure}

\Cref{quantization} shows the sum-rate when the discretization stage is applied to the \ac{IRS} phase-shift matrix obtained for a configuration with $K=10$, $N_t = 2$, $N_s=2$, $N_r=30$, $N=50$ and SNR $=10$ dB. We use the continuous-valued \ac{IRS} phase-shift matrix obtained by the \ac{DCB-DDPG} framework as a baseline. The figure shows that the sum-rate degradation caused by discretizing the \ac{IRS} phase-shifts is lower than 20 bit/s/Hz, even considering only two possible phases for the \ac{IRS} scattering elements. Besides, this degradation effect steeply decreases when more phase-shift quantization levels are considered. As observed, we reach almost the same performance with only eight quantization levels. Comparable results have been obtained for other system configurations. In addition, a similar codebook-based approach could be considered for the precoders. Hence, continuous-valued formulations are a good choice for the system design, even if discrete-valued matrices are the final objective.

\section{Conclusions}
We have investigated two different approaches for the joint optimization of the IRS phase-shift matrix and the precoders in the uplink of an IRS-assisted MS MU-MIMO system. The first approach is a \ac{CB} formulation with continuous state and action spaces. To handle this continuous formulation, we have developed an actor-critic framework called DCB-DDPG. The second approach is a \ac{DRL} formulation of the optimization problem. By considering the pros and cons of related works, we developed an approach to the DDPG framework with an alternative configuration of the state and action spaces, such that entries in the current matrices are the states and the variations in these values are the actions. A deep analysis of both proposed frameworks has been made, and our main findings can be summarized as follows:
\begin{itemize}
\item{CB and RL enable an adequate formulation of the joint optimization, while DCB-DDPG and DRL-DDPG frameworks are effective methods to solve the considered optimization problem.}

\item{The convergence of both proposed frameworks in terms of reward maximization and critic loss minimization is empirically illustrated. However, DCB-DDPG shows better stability and convergence speed, which can be related to its lower computational complexity and better suitability of the problem formulation.}

\item {The performance of both proposed frameworks regarding sum-rate maximization is similar to state-of-the-art heuristic algorithms when $N_\text{r}\geq KN_\text{s}$. In contrast, the proposed schemes handle the multi-user interference more efficiently when $N_\text{r}<KN_\text{s}$.}

\item {Simulation results show that our continuous-valued formulations are valuable even when discrete-valued matrices are required as final outcomes.} 

\end{itemize}

\section*{Acknowledgments}
This work has been supported by grants ED431C 2020/15 and ED431G 2019/01 (to support the Centro de Investigación de Galicia “CITIC”) funded by Xunta de Galicia and ERDF Galicia 2014-2020; and by grants PID2019-104958RB-C42 (ADELE) and BES-2017-081955 funded by MCIN/AEI/10.13039/501100011033.
\bibliographystyle{IEEEtranTCOM}
\bibliography{IEEEabrv,bibtex/IRS_DRL}

\vfill

\end{document}